\documentclass[useAMS,usenatbib]{mnras}

\usepackage{natbib}
\usepackage{epsfig} 
\usepackage{scrtime}
\usepackage{graphicx}	
\usepackage{amsmath}	
\usepackage{amssymb}	
\usepackage{float}
\usepackage[caption = false]{subfig}
\usepackage{multirow}
\usepackage{rotating}
\usepackage{parskip}

\newcommand{\logd}{\log}

\newcommand{\mincir}{\raise-3.truept\hbox{\rlap{\hbox{$\sim$}}\raise4.truept\hbox{$<$}\ }}
\def\mean#1{\left< #1 \right>}

\title[2D Surface Brightness of Large 2MASS Galaxies I]{2D Surface Brightness Modelling of Large 2MASS Galaxies I: Photometry and Structural Parameters}
\author[R\'ios-L\'opez et al.]{Emmanuel R\'ios-L\'opez,$^{1,2}$\thanks{(E-mail:riloemm@inaoep.mx)}
Christopher A\~norve,$^{2}$\thanks{(E-mail:canorve@uas.edu.mx)} H\'ector J. Ibarra-Medel,$^{1,3}$ 
\newauthor
Omar L\'opez-Cruz,$^{1}$
Joaqu{\'\i}n Alvira-Enr{\'\i}quez,$^{4}$
Gabriela Iacobuta$,^{5,6,7}$
Mabel Valerdi$^{1,8}$
\\
$^{1}$Instituto Nacional de Astrof\'isica, \'Optica y Electr\'onica, Tonantzintla, Puebla, M\'exico \\
$^{2}$Facultad de Ciencias de la Tierra y del Espacio, Universidad Aut\'onoma de Sinaloa, Culiac\'an, Sinaloa, M\'exico\\
$^{3}$University of Illinois Urbana-Champaign, Department of Astronomy, 1002 W. Green St, Urbana, Illinois, 61801, United States\\
$^{4}$Universidad Iberoamericana, Department of Physics, Mexico City,  Mexico\\
$^{5}$School of Physics and Astronomy, University of Nottingham, Nottingham, NG7 2RD, UK \\
$^{6}$German Development Institute/Deutsches Institut fur Entwicklungspolitik (GDI/DIE), Tulpenfeld 6, Bonn \\
$^{7}$Environmental Systems Analysis Group, Wageningen University \& Research, PO Box 47,
Wageningen 6700 AA, The Netherlands\\
$^{8}$Instituto de Astronom{\'\i}a, Universidad Nacional Aut\'onoma de M\'exico, Av. Universidad 3000, Ciudad Universitaria, C.P. 04510,
M\'exico
}

\pubyear{2021}

\begin{document}
\date{Accepted XXX. Received YYY; in original form ZZZ}
\pagerange{\pageref{firstpage}--\pageref{lastpage}}
\maketitle
\label{firstpage}

\begin{abstract}

We have studied a sample of 101 bright 2MASS galaxies from the Large Galaxy Atlas (LGA), whose morphologies span from early to late-types. We have generated estimates for structural parameters through a two-dimensional (2D) surface brightness photometric decomposition in the three 2MASS bands ($J,H,K_s$). This work represents a detailed  multi-component photometric study of nearby galaxies. We report total magnitudes, effective radii, concentration indices, among other parameters, in the three 2MASS bands. We found that the integrated total magnitudes of early-type galaxies (ETGs) measured on 2MASS LGA mosaics are $\sim$0.35 mag dimmer, when compared with images generated from IRSA image tiles service; nevertheless, when comparing late-type galaxies (LTGs) we did not find any difference. Therefore, for ETGs we present the results derived on IRSA image tiles, while for LTGs we used data from the LGA mosaics. Additionally, by combining these structural parameters with scaling relations and kinematic data, we separated classical bulges from pseudobulges. We found that $\sim$40$\%$ of the objects in our sample are classified as pseudobulges, which are found preferentially in LTGs. Also, our findings confirm trends reported earlier in the distributions for some physical parameters, such as S\'ersic index, $B/T$ and $q$ ratios. In general, our results are in agreement with previous one-dimensional studies. In a companion paper, we revise some of the scaling relations among global galaxy properties, as well as their interrelation with Supermassive Black Holes.
\vspace{.10cm}
\end{abstract}


\section{Introduction}\label{Intro}
The generally accepted cold dark matter scenario (CDM) is based on a hierarchical galaxy formation model \citep[e.g.,][]{White-Rees_1978}. If dark energy is included, then we have the general $\Lambda$CDM framework \citep[see, e.g.,][]{Springel_et_al_2005, Spergel_et_al_2007}, which satisfactorily explains galaxy formation and evolution. Nevertheless, complementary physical processes must be included through semi-analytical approaches \citep[e.g.,][]{Silk_et_al_2015} to have a comprehensive view of such mechanisms.

The morphological classification of galaxies suggests the presence of a common process which takes place during galaxy formation and evolution. But, the photometric structural analysis of galaxies provides qualitative information that supports galaxy classification schemes, which also helps to constrain the dynamical modelling of galaxies \citep[e.g.,][]{1990ApJ...361...78B,1991MNRAS.253..710V,2013MNRAS.432.1709C}. Moreover, \citet{2020arXiv200507588P} suggested that the distribution of the bulge-to-disc ratio in nearby galaxies, can help us to constrain cosmological initial conditions. Hence, the accurate modelling of the surface brightness of galaxies acquires additional relevance. 

The space of parameters relating dynamics, surface brightness and scale size can be reduced to conform the Fundamental Plane \citep[FP,][]{Djorgovski-Davis_1987}. Some previously established  correlations such as the Faber-Jackson \citep{Faber-Jackson_1976}, Kormendy \citep{Kormendy_1977} and the Tully-Fisher \citep{Tully-Fisher_1977}, could be considered projections of the FP. These correlations might suggest that other connections among galaxy properties might be present.

Observations of nearby galaxies allow us to attain large angular resolution, which in turn allows us to analyse the distribution of internal structures such as bulges, spiral arms, bars, rings, among others. The study of those structures in large galaxy samples has helped us to learn that smaller bulges are similar to low-mass compact elliptical galaxies in the local Universe, while larger bulges are equivalent to the massive compact galaxies in the distant Universe \citep[e.g.,][]{Dullo-Graham_2013}.

\begin{figure*}
\begin{center}
\includegraphics[width=17.65 cm]{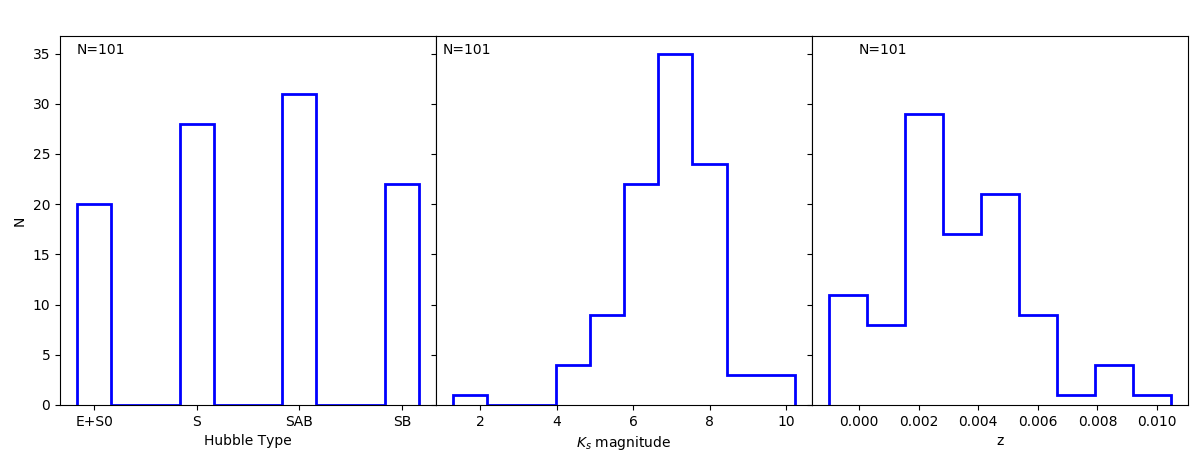}
\caption[Hist-Hubble_type]{\footnotesize Sample distribution of our galaxies in Hubble Type (left panel), brightness in $K_s$ band (middle panel) and redshift (right panel).}
\label{Hist-Hubble_type}
\end{center}
\end{figure*}

As stated above, bulges have become very relevant due to their linkages with the different mechanisms of formation and evolution of galaxies \citep[e.g.,][]{Athanassoula_2005}. Bulges can be divided into two classes established observationally: classical bulges and pseudobulges \citep{Kormendy_et_al_2011}. Classical bulges resemble E galaxies and share some physical aspects such as that they are systems dominated by velocity dispersions, have a higher S\'ersic index (usually above 2) and are populated by old stars. Besides, classical bulges follow the same scaling relations traced by ellipticals. This tells us about their formation history, which is related also to the merger scenario as early-type galaxies (ETGs). On the other hand, pseudobulges are more disc-like systems and different from classical and E galaxies. Thus, pseudobulges are systems that display more flattening as discs, as well as they are dominated by rotation velocities, tend to have lower S\'ersic indices and young stars are the dominant population. The formation of pseudobulges is attributed to secular processes \citep[see][for reviews about properties of bulge types]{Kormendy-Kennicutt_2004, Kormendy-Ho_2013}.

We have employed a two-dimensional (2D) approach to modelling the light distribution of 101 galaxies observed in the near-infrared (NIR), while many studies have been based on one-dimensional (1D) surface photometry to estimate structural parameters \citep[e.g.,][and references therein]{Peng_et_al_2010, Schombert_2012}. 1D methods, like isophotal analysis, are suitable for working in poor signal-to-noise ratio (S/N) conditions, since the points on the light profile are the result of the azimuthally averaged surface brightness along the ellipse, as well as the fact that the additional components for each isophote can be known in a relatively direct and easy way \citep[e.g.,][]{Jedrzejewski_1987}. A drawback of this procedure is that in the presence of multiple galaxy components with different orientations, the selection of major or minor axes could be inaccurate. On the other hand, 2D decomposition methods have the advantage that are able to disentangle among galaxy components, such as bulges, discs and bars, breaking part of the degeneracy in the parameters due to the fact that those components may have different ellipticities and position angles. In addition, the 2D approach can provide superior capabilities by convolving the model with the point-spread function (PSF). However, a weakness in this case is related to the fact that perfect ellipsoid models may not be entirely appropriate for describing those galaxies with isophotal twists. The differences between 1D and 2D techniques have been widely discussed in the literature \citep[e.g.,][and references therein]{Ravindranath_et_al_2001,Peng_et_al_2010, Bonfini_2014,2017ApJ...845..114G}. 

Some previous photometric studies have been limited to only two components: discs and bulges, disregarding bars \citep[e.g.,][]{2011ApJS..196...11S}. Not including bars may alter the outcome of the decomposition (mainly in magnitude, effective radius and S\'ersic index), related to the overestimation and uncertainties of such values \citep[e.g.,][]{2008MNRAS.384..420G,Laurikainen_et_al_2004_2,Fisher-Drory_2008,2017A&A...598A..32M}. Other studies have  held some parameters fixed during the fit, for instance the S\'ersic index $n$ \citep[e.g.,][]{Bruce_et_al_2012}. This practice affects the estimation of parameter and their errors, as well as the convergence of fits \citep[e.g.,][]{Peng_et_al_2010}. Nevertheless, in the optical there have been some recent 2D studies which include multiple components overcoming the difficulties mentioned above, specifically  the work by the group associated to Carnegie-Irvine Galaxy Survey \citep[CGS; e.g.,][]{2013ApJ...766...47H,2017ApJ...845..114G,2019ApJS..244...34G}. Complementary to the CGS results, we present the results of an unrestricted 2D multi-component NIR photometric study of nearby galaxies, using data from Two Micron All Sky Survey \citep[2MASS,][]{2006AJ....131.1163S}. Our sample includes early-type (E and S0) and late-type (S, SAB and SB) galaxies. We have used GALFIT \citep{Peng_et_al_2002, Peng_et_al_2010}, allowing three initial components:  bulges, discs and bars, in some cases an additional PSF component was considered when dealing with active galactic nuclei (AGN). A comprehensive discussion on the applicability of different models is presented.

Surface brightness studies conducted in the NIR are less affected by gas and dust extinction. Moreover, the galaxies' light in the NIR is dominated by the older stellar populations, which represent the main baryonic component in most galaxies. Hence, NIR observations enhance the contrast between the underlying mass component (older stellar population) and the younger stellar population components, whose light dominates in the optical bands \citep[e.g.,][]{Frogel_et_al_1996,1993ApJ...418..123R,1996A&AS..118..557D,1996A&A...313..377D,Jarrett_et_al_2003}.

The paper is organised as follows: \S \ref{Data} describes the sample and the NIR observations from 2MASS survey, as well as additional data used in this work. In \S \ref{Met} we present our methodology, mainly the 2D multi-component photometric decomposition performed with GALFIT. \S \ref{Res} presents the results obtained from surface brightness modelling of the galaxies in the NIR bands of 2MASS, as well as the classification performed to separate bulge types. In \S\ref{Dis} we discuss our findings. We present our conclusions in \S \ref{Con}. We also have explored some galaxy scaling relations using the physical properties derived here, along with kinematic data taken from the literature, as well as those relations between host galaxy properties and their Supermassive Black Holes (SMBHs). Such results are presented and discussed in companion paper (R\'ios-L\'opez et al. 2021, hereafter Paper II). Unless stated otherwise, we have adopted $\Omega_m=0.3$,  $\Omega_{\Lambda}=0.7$ and $\mathrm{H_0}=70\,h_{70}\;  \mathrm{km\,s^{-1}\, Mpc^{-1}}$ throughout the paper.

\section{Data}\label{Data}

\subsection{Sample and Observations}\label{Sample_Obs}

Table \ref{Table_sample} presents the sample considered in this study. It contains 101 galaxies, which is a subsample of the 2MASS \citep{2006AJ....131.1163S} Large Galaxy Atlas\footnote{LGA data available at \url{https://irsa.ipac.caltech.edu/applications/2MASS/LGA/atlas.html}} \citep[LGA,][]{Jarrett_et_al_2003}. We have selected some of the nearest ($z\leq 0.01$) and brightest ($K_{s} \leq 10$) sources in the 2MASS survey \citep{Jarrett_2004}. In Figure \ref{Hist-Hubble_type} we present the distribution according to morphological type, magnitude in $K_s$ band and redshift\footnote{Redshift data from the NASA Extragalactic Database at \url{http://ned.ipac.caltech.edu/}}. Furthermore, about half of our objects are part of the 100 largest galaxies in angular size observed with 2MASS \citep{Jarrett_et_al_2003}, some of them  can also be found in the  ``Atlas de Galaxias Australes'' by Jos\'e Luis \citet{Sersic_1968}.

The 2MASS survey began operations in the mid-1990s, completing its  observations by early 2001. 2MASS used twin 1.3-meter aperture telescopes located in both equatorial hemispheres (the northern telescope at the summit of Mt. Hopkins, Arizona, while the southern telescope at the summit of Cerro Tololo, Chile) to survey  the whole sky, detecting more than 500 million stars (Point Source Catalog, PSC) and resolving more than 1.5 million galaxies (Extended Source Catalog, XSC) in the NIR bands \citep{Jarrett_2004}. 2MASS observed the sky in a drift-scan mode with $8\farcm5\,\times\, 6\degr$  tiles or ``scans'', forming 23 separate images per tile per band of $8\farcm5 \times 17\arcmin$ coadds. Each image is obtained from six pointings with a total integration time of $t_{int}=7.8 \; \mathrm{s/pixel}$, with a plate scale of $1\arcsec/\mathrm{pixel}$ \citep[for more details, see][]{Skrutskie_et_al_1997,2006AJ....131.1163S}.  

Particularly, for the LGA  a set of custom-made mosaics and additional data products  were generated for more than 600 objects, classified along the Hubble type: elliptical, S0s, spirals and barred spirals, as well as dwarf and peculiar galaxies. Galaxies range in size from $2\arcmin$ to $2\degr$, with a spatial resolution (PSF FWHM, depending on the atmospheric seeing) of $\sim$2$\arcsec$-3$\arcsec$ in the $J\; (1.2\, \mathrm{\mu m})$, $H \; (1.6\, \mathrm{\mu m})$ and $K_{s}\; \mathrm{(2.2\, \mu m})$ bands. LGA images reach a $1\;\sigma$ background noise at $J=21.4\;\mathrm {mag/arcsec^2}$, $H=20.6 \;\mathrm {mag/arcsec^2}$ and $K_s = 20.0 \;\mathrm {mag/arcsec^2}$ \citep{Jarrett_et_al_2003}. Therefore, LGA sample has enough sensitivity and angular resolution to allow  detailed studies of galaxy structure. These images show spiral arms, bulges, bars and star forming regions \citep[for more details, see][and references therein]{Jarrett_et_al_2003}. In addition, NIR observations, as mentioned in \S\ref{Intro}, are less affected by extinction than in the optical. Nevertheless, it was brought to our attention that LGA mosaics suffered from background over-subtraction \citep[Jarrett 2011, private communication cited in][]{Schombert_2012} produced by using not-sufficiently large apertures around bright ETGs. This produced an underestimation of the background during the mosaic generation\footnote{See the discussion by Tom Jarret in: \url{https://wise2.ipac.caltech.edu/staff/jarrett/2mass/ellipticals.html}}. As expected, a deficit in the total magnitudes of ETGs was found \citep[e.g.,][]{Schombert_2012}, causing inaccuracies in other related parameters, such as effective radii. The resulting background over-subtraction error on the LGA mosaics has some affinity with the one reported by \citet{2007ApJ...662..808L} for SDSS photometry data on ETGs, which was corrected in the later SDSS data releases \citep[e.g.,][for DR8]{2011AJ....142...31B}. Below (see \S\ref{LGA_VS_tile}), we present a detailed analysis using LGA mosaics and 2MASS image tiles downloaded from the NASA/IPAC Infrared Science Archive (IRSA)\footnote{2MASS image tiles service \url{https://irsa.ipac.caltech.edu/applications/2MASS/IM/interactive.html##pos}}. 

According to their 2MASS $K_{s}$-band integrated fluxes and distances (see Figure \ref{Hist-Hubble_type}), galaxies in Table \ref{Table_sample} are among the brightest nearby quiescent galaxies (columns 4 to 7 in Table \ref{Table_sample}). We have directed our selection to cover the distribution of morphological types, from early to late type systems, such as ellipticals, lenticulars, intermediate and barred spirals (column 2 in Table \ref{Table_sample}); besides, by covering most of Hubble types and especially disc galaxies, we are able to examine internal structures over a wide range of luminosities and sizes, allowing the exploration of classical bulges and pseudobulges distribution.

\subsection{Velocity Dispersion Data}

Stellar velocity dispersion data \textbf{(column 8 in Table \ref{Table_sample})} were taken from the literature, mainly from the work of \citet{Ho_et_al_2009} and Hyperleda database\footnote{Data available at \url{http://leda.univ-lyon1.fr/leda/param/vdis.html}} \citep{Paturel_et_al_2003}.

\begin{figure*}
\begin{center}
\includegraphics[width=17.65 cm]{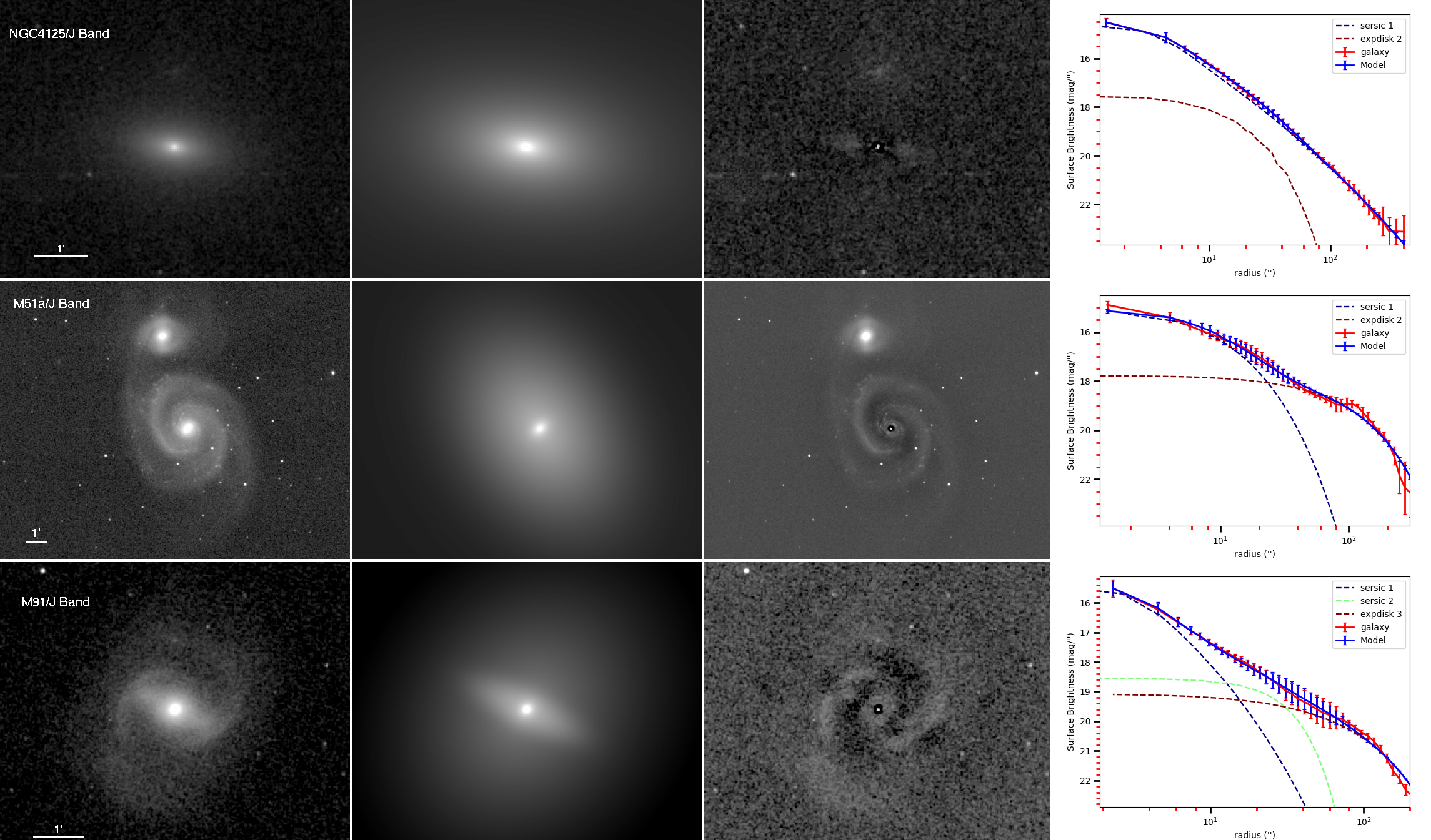}
\caption[varias_3bands]{\footnotesize Examples of GALFIT decomposition method for the galaxies NGC4125, M51a, and M91 from top to bottom, respectively. The images are, from left to right: 2MASS image, model and residual images generated with GALFIT and 1D surface brightness profile generated with EllipSect.}
\label{varias_3bands}
\end{center}
\end{figure*}

\begin{table*}
	\centering
	\caption{Sample of Large 2MASS Galaxies.}
	\label{Table_sample}
	\begin{tabular}{lcccllll}
	\hline
Name & Hubble Type & $v_{h}$ & Dist. & $K_{s}$ & $K_{s}$ & $M_{K_s}$  & $\sigma$\\
\multicolumn{2}{c}{}&\multicolumn{1}{l}{$[\mathrm{km\,s}^{-1}]$} & [Mpc] & [2MASS] & & & $[\mathrm{km\,s}^{-1}]$\\
(1) & (2) & (3) &  (4) & (5) & (6) & (7) & (8) \\ \hline
\multicolumn{8}{c}{------E+S0 Galaxies------} \\
M110		 & 	E5;pec	 & 	-241 $\pm$ 3	& 0.82  &  5.59 $\pm$ 0.04 & 4.67 $\pm$ 0.06 & -19.90 $\pm$ 0.06 & 23 $\pm$ 4 [a]\\
M32	 & 	cE2		 & 	-213 $\pm$ 2	 & 0.80  & 	5.10 $\pm$ 0.02 & 5.08 $\pm$ 0.05 & -19.45 $\pm$ 0.05 & 72 $\pm$ 2 [a]\\
Maffei1	 & S0-pec	 & 	66 $\pm$ 5 & 2.85 & 4.68 $\pm$ 0.02 & 4.43 $\pm$ 0.07 & -22.84 $\pm$ 0.07 & 187 $\pm$ 6 [b]\\
NGC1549	 & 	E0-1 & 	1256 $\pm$ 12 & 16.63 & 6.78 $\pm$ 0.02 & 6.51 $\pm$ 0.06 & -24.59 $\pm$ 0.06 & 199 $\pm$ 4 [b] \\
NGC1947	 & 	S0-pec	 & 	1100 $\pm$ 24 & 16.30  & 7.51 $\pm$ 0.03 & 6.86 $\pm$ 0.06 & -24.20 $\pm$ 0.06 & 173 $\pm$ 12 [a]\\
NGC2768	 & 	S0;1/2 & 1353 $\pm$ 5 & 20.46 & 7.00 $\pm$ 0.03 & 6.80 $\pm$ 0.06 & -24.76 $\pm$ 0.06 & 182 $\pm$ 4 [a]\\
NGC3115	 & 	S0-	 & 	663 $\pm$ 4 & 9.54 & 5.88 $\pm$ 0.02 & 5.67 $\pm$ 0.05 & -24.22 $\pm$ 0.05 & 252 $\pm$ 6 [a]\\
NGC3377	 & 	E5-6  & 665 $\pm$ 2 & 10.99 & 7.44 $\pm$ 0.03 & 7.26 $\pm$ 0.09 & -22.94 $\pm$ 0.09 & 139 $\pm$ 3 [a]\\
M105	 & 	E1	 & 	911 $\pm$ 2 & 10.70 & 6.27 $\pm$ 0.02 & 6.23 $\pm$ 0.04 & -23.92 $\pm$ 0.04 & 207 $\pm$ 2 [a] \\
NGC4125	 & 	E6;pec	 & 	1281 $\pm$ 14 & 22.76 & 6.86 $\pm$ 0.02 & 5.87 $\pm$ 0.07 & -25.92 $\pm$ 0.07 & 227 $\pm$ 8 [a]\\
NGC4365	 & E3 & 	1243 $\pm$ 5 & 21.62  & 6.64 $\pm$ 0.03 & 6.26 $\pm$ 0.07 & -25.41 $\pm$ 0.07 & 256 $\pm$ 3[a] \\
M86		 & 	S0(3)/E3	 & 	-224 $\pm$ 5 & 16.40 &  6.10 $\pm$ 0.03 & 5.18 $\pm$ 0.06 & -25.89 $\pm$ 0.06 & 235 $\pm$ 3 [a] \\
M49	 	 & 	E2/S0(2) & 	981 $\pm$ 5 & 16.72 & 5.40 $\pm$ 0.02 & 5.36 $\pm$ 0.04 & -25.76 $\pm$ 0.04 & 291 $\pm$ 3 [a]\\
M87		 & 	E+0-1pec	 & 	1284 $\pm$ 5 & 16.68 & 5.81 $\pm$ 0.02 & 5.66 $\pm$ 0.05 & -25.45 $\pm$ 0.05 & 332 $\pm$ 5 [a] \\
NGC4636	 & 	E/S0$_1$	 & 	938 $\pm$ 4 & 13.70 & 6.42 $\pm$ 0.04 & 5.46 $\pm$ 0.05 & -25.22 $\pm$ 0.05 & 203 $\pm$ 3 [a]\\
M60		 &	E2 &  1110 $\pm$ 5 & 16.46 & 5.74 $\pm$ 0.02 & 5.25 $\pm$ 0.05 & -25.83 $\pm$ 0.05 & 335 $\pm$ 4 [a]\\
NGC4697	 & 	E6		 & 1241 $\pm$ 2	& 12.54 & 6.37 $\pm$ 0.03 & 5.91 $\pm$ 0.07 & -24.58 $\pm$ 0.07 & 172 $\pm$ 6 [b] \\
NGC4976	 & 	E4;pec	 & 	1453 $\pm$ 24 & 12.18 & 6.85 $\pm$ 0.02 & 6.23 $\pm$ 0.06 & -24.20 $\pm$ 0.06 & 139 $\pm$ 13 [b]\\
NGC5084	 & 	S0		 & 	1721 $\pm$ 3 & 41.37 & 7.06 $\pm$ 0.03 & 6.93 $\pm$ 0.05 & -26.15 $\pm$ 0.05 & 201 $\pm$ 6 [b]\\
IC5328	 & 	E		 & 	3137 $\pm$ 13 & 37.80 & 8.28 $\pm$ 0.03 & 8.08 $\pm$ 0.12 & -24.81 $\pm$ 0.12 & 206 $\pm$ 8 [b]\\
\multicolumn{8}{c}{------S Galaxies------}\\
M31	 & 	SA(s)b	 & -301 $\pm$ 1 & 0.77 & 0.98 $\pm$ 0.02 & 0.87 $\pm$ 0.04 & -23.58 $\pm$ 0.04 & 170 $\pm$ 5 [a] \\
M33	 & 	SA(s)cd	 & 	-180 $\pm$ 1 & 0.82 & 4.10 $\pm$ 0.04 & 3.94 $\pm$ 0.06 & -20.63 $\pm$ 0.06 & 21 $\pm$ 3 [a] \\
NGC1553	 & 	SA(rl)0	 & 1080 $\pm$ 11 & 18.40 & 6.28 $\pm$ 0.02 & 6.26 $\pm$ 0.04 & -25.06 $\pm$ 0.04 & 186 $\pm$ 4 [b] \\
NGC2541	 & 	SA(s)cd	 & 	548 $\pm$ 1 & 11.50 &   10.09 $\pm$ 0.05 & 9.37 $\pm$ 0.04 & -20.93 $\pm$ 0.04 & 53 $\pm$ 10 [a] \\
NGC2683	 & 	SA(rs)b	 & 411 $\pm$ 1 & 11.70 &   6.33 $\pm$ 0.02 & 6.23 $\pm$ 0.02 & -24.11 $\pm$ 0.02 & 130 $\pm$ 7[a] \\
NGC2775	 & 	SA(r)ab	 & 1350 $\pm$ 2 & 17.00 &   7.06 $\pm$ 0.02 & 7.01 $\pm$ 0.05 & -24.14 $\pm$ 0.05 & 174 $\pm$ 8 [a] \\
NGC2985	 & 	SA(rs)ab	 & 1324 $\pm$ 1 & 20.60 &   7.36 $\pm$ 0.03 & 7.34 $\pm$ 0.03 & -24.23 $\pm$ 0.03 & 141 $\pm$ 5 [a] \\
M81	     & 	SA(s)ab;Sy & -39 $\pm$ 3 & 3.60 &   3.83 $\pm$ 0.02 & 3.80 $\pm$ 0.03 & -23.98 $\pm$ 0.03 & 162 $\pm$ 3 [a] \\
NGC3631	 & 	SA(s)c	 &	1156 $\pm$ 1 & 8.69 &   7.98 $\pm$ 0.06 & 7.70 $\pm$ 0.05 & -21.99 $\pm$ 0.05 & 44 $\pm$ 9 [a] \\
NGC3675	 & 	SA(s)b	 & 770 $\pm$ 1 & 12.40 &   6.86 $\pm$ 0.02 & 6.83 $\pm$ 0.05 & -23.63 $\pm$ 0.05 & 108 $\pm$ 4 [a] \\
NGC3877	 & 	Sc		 & 895 $\pm$ 1 & 15.20 &   7.75 $\pm$ 0.02 & 7.68 $\pm$ 0.05 & -23.23 $\pm$ 0.05 & 86 $\pm$ 9 [a] \\
NGC3938	 & 	SA(s)c	 & 808 $\pm$ 2 & 17.90 &   7.81 $\pm$ 0.05 & 7.79 $\pm$ 0.06 & -23.48 $\pm$ 0.06 & 29 $\pm$ 5 [a] \\
NGC4013	 & 	SAb		 & 831 $\pm$ 1 & 17.80 &   7.63 $\pm$ 0.02 & 7.58 $\pm$ 0.04 & -23.67 $\pm$ 0.04 & 86 $\pm$ 9 [a] \\
NGC4254	 & 	SA(s)c	 & 2406 $\pm$ 1 & 14.30 &   6.93 $\pm$ 0.03 & 6.72 $\pm$ 0.04 & -24.06 $\pm$ 0.04 & 83 $\pm$ 9 [a] \\
M85	     & 	SA(s)0;pec & 729 $\pm$ 2 & 17.90 &   6.14 $\pm$ 0.02 & 6.12 $\pm$ 0.04 & -25.16 $\pm$ 0.04 & 179 $\pm$ 5 [a] \\
M88     & 	SA(rs)b	 & 2284 $\pm$ 1 & 16.50 &   6.27 $\pm$ 0.02 & 6.26 $\pm$ 0.05 & -24.83 $\pm$ 0.05 & 167 $\pm$ 7 [a] \\
M104     & 	SA(s)a;Sy	 & 1024 $\pm$ 5 & 9.87 &   4.96 $\pm$ 0.02 & 4.95 $\pm$ 0.02 & -25.13 $\pm$ 0.02 & 241 $\pm$ 4 [a] \\
NGC4710	 & 	SA(r)0+	 & 1102 $\pm$ 5 & 16.80 &   7.57 $\pm$ 0.02 & 7.55 $\pm$ 0.04 & -23.81 $\pm$ 0.04 & 110 $\pm$ 10 [a] \\
NGC4826	 & 	SA(rs)ab;Sy & 409 $\pm$ 1 & 7.27 &   5.33 $\pm$ 0.02 & 5.31 $\pm$ 0.04 & -24.00 $\pm$ 0.04 & 96 $\pm$ 3 [a] \\
NGC4866	 & 	SA(r)0+sp	 & 1980 $\pm$ 3 & 31.10 &   7.92 $\pm$ 0.02 & 5.97 $\pm$ 0.06 & -24.80 $\pm$ 0.06 & 210 $\pm$ 8 [a] \\
NGC5033	 & 	SA(s)c	 & 875 $\pm$ 1 & 19.30  &   6.96 $\pm$ 0.03 & 6.94 $\pm$ 0.02 & -24.49 $\pm$ 0.02 & 151 $\pm$ 4 [a] \\
M63	     &  SA(rs)bc	 & 500 $\pm$ 1 & 8.90 &  5.61 $\pm$ 0.02 & 5.55 $\pm$ 0.03 & -24.20 $\pm$ 0.03 & 117 $\pm$6 [a] \\
NGC5102	 & 	SA0-		 & 468 $\pm$ 2 & 3.20 &  6.92 $\pm$ 0.04 & 6.90 $\pm$ 0.05 & -20.64 $\pm$ 0.05 & 66 $\pm$ 10 [b] \\
M51a     & 	SA(s)bc	 & 460 $\pm$ 2 & 7.90 &  5.50 $\pm$ 0.02 & 5.35 $\pm$ 0.05 & -24.13 $\pm$ 0.05 & 96 $\pm$ 9 [a] \\
NGC5317	 & 	SA(rs)bc;pec & 1268 $\pm$ 2 & 13.60 & 7.80 $\pm$ 0.05 & 7.69 $\pm$ 0.11 & -22.97 $\pm$ 0.11 & 22 $\pm$ 9 [b] \\
NGC6015	 & 	SA(s)cd	 &	833 $\pm$ 1 & 18.90 &   8.47 $\pm$ 0.04 & 8.40 $\pm$ 0.06 & -22.98 $\pm$ 0.06 & 44 $\pm$ 9 [a] \\
NGC6503	 & 	SA(s)cd	 & 	25 $\pm$ 1 & 5.30 & 7.30 $\pm$ 0.02 & 7.28 $\pm$ 0.05 & -21.34 $\pm$ 0.05 & 46 $\pm$ 3 [a] \\
NGC7793	 & 	SA(s)d	 & 227 $\pm$ 2 & 3.40 &   6.86 $\pm$ 0.06 & 6.80 $\pm$ 0.06 & -20.85 $\pm$ 0.06 & $\cdots$ \\
\hline
\end{tabular}
\end{table*}

\begin{table*}
	\centering
	\begin{tabular}{lcccllll}
	\hline
Name & Hubble Type & $v_{h}$ & Dist. & $K_{s}$ & $K_{s}$ & $M_{K_s}$ & $\sigma$\\
\multicolumn{2}{c}{}&\multicolumn{1}{l}{$[\mathrm{km\,s}^{-1}]$} & [Mpc] & [2MASS] & & & $[\mathrm{km\,s}^{-1}]$\\
(1) & (2) & (3) &  (4) & (5) & (6) & (7) & (8) \\ \hline
\multicolumn{8}{c}{------SAB Galaxies------} \\
NGC253	 &  SAB(s)c	 & 242 $\pm$ 1 & 3.50 & 3.77 $\pm$ 0.02 & 3.76 $\pm$ 0.05 & -23.97 $\pm$ 0.05 & 109 $\pm$ 20 [b] \\
NGC1316	 & 	SAB(s)0	 & -107 $\pm$ 8 & 20.95 & 5.59 $\pm$ 0.02 & 5.48 $\pm$ 0.02 & -26.12 $\pm$ 0.02 & 219 $\pm$ 10 [b] \\
IC342	 & 	SAB(rs)cd	 & 31 $\pm$ 3 & 3.73 & 4.56 $\pm$ 0.04 & 4.38 $\pm$ 0.05 & -23.47 $\pm$ 0.05 & 74 $\pm$ 11 [a] \\
NGC2403	 & 	SAB(s)cd	 & 133 $\pm$ 1 & 3.06 & 6.19 $\pm$ 0.04 & 6.16 $\pm$ 0.06 & -21.27 $\pm$ 0.06 & 68 $\pm$ 32 [a] \\
NGC2715	 & 	SAB(rs)c	 & 1323 $\pm$ 1 & 16.40 & 8.60 $\pm$ 0.04 & 8.59 $\pm$ 0.30 & -22.49 $\pm$ 0.30 & 85 $\pm$ 9 [a] \\
NGC3166	 & 	SAB(rs)0/a	 & 1183 $\pm$ 1 & 22.00 & 7.21 $\pm$ 0.02 & 7.17 $\pm$ 0.03 & -24.54 $\pm$ 0.03 & 153 $\pm$ 8 [a] \\
NGC3184	 &  SAB(rs)cd	 & 582 $\pm$ 1 & 9.68 & 7.22 $\pm$ 0.07 & 7.03 $\pm$ 0.06 & -22.90 $\pm$ 0.06 & 43 $\pm$  9 [a] \\
NGC3344	 & 	SAB(r)bc	 & 580 $\pm$ 1 & 8.28 & 7.44 $\pm$ 0.04 & 7.39 $\pm$ 0.05 & -22.20 $\pm$ 0.05 & 74 $\pm$ 9 [a] \\
M96		 & 	SAB(rs)ab  & 888 $\pm$ 1 & 10.62 & 6.32 $\pm$ 0.02 & 6.33 $\pm$ 0.04 & -23.80 $\pm$ 0.04 & 128 $\pm$ 4 [a] \\
NGC3486	 & 	SAB(r)c	 & 678 $\pm$ 1 & 12.60 & 8.00 $\pm$ 0.04 & 8.09 $\pm$ 0.04 & -22.41 $\pm$ 0.04 & 65 $\pm$ 3 [a] \\
M65		 & 	SAB(rs)a	 & 803 $\pm$ 2 & 14.60 & 6.07 $\pm$ 0.02 & 6.00 $\pm$ 0.04 & -24.82 $\pm$ 0.04 & 138 $\pm$ 3 [a] \\
NGC3726	 & 	SAB(r)c	 & 864 $\pm$ 1 & 13.00 & 7.78 $\pm$ 0.05 & 7.61 $\pm$ 0.06 & -22.96 $\pm$ 0.06 & 42 $\pm$ 9 [a] \\
NGC4157	 & 	SAB(s)b	 & 771 $\pm$ 1 & 15.10 & 7.36 $\pm$ 0.02 & 7.31 $\pm$ 0.02 & -23.89 $\pm$ 0.02 & 90 $\pm$ 4 [a] \\
M98     & 	SAB(s)ab	 & -142 $\pm$ 4 & 13.60 & 6.89 $\pm$ 0.02 & 6.83 $\pm$ 0.05 &  -23.83 $\pm$ 0.05 & 132 $\pm$ 7 [a] \\
NGC4216	 & 	SAB(s)b	 & 131 $\pm$ 4 & 14.10 & 6.52 $\pm$ 0.02 & 6.50 $\pm$ 0.04 & -24.25 $\pm$ 0.04 & 197 $\pm$ 8 [a] \\
M106	 & 	SAB(s)bc	 & 448 $\pm$ 3 & 7.27 & 5.46 $\pm$ 0.02 & 5.43 $\pm$ 0.03 & -23.88 $\pm$ 0.03 & 148 $\pm$ 4 [a] \\
M61		 & 	SAB(rs)bc	 & 1566 $\pm$ 2 & 12.30 & 6.84 $\pm$ 0.03 & 6.74 $\pm$ 0.04 & -23.71 $\pm$ 0.04 & 84 $\pm$ 3 [a] \\
M100	 & 	SAB(s)bc	 & 1571 $\pm$ 1 & 14.20 & 6.59 $\pm$ 0.04 & 6.50 $\pm$ 0.03 & -24.26 $\pm$ 0.03 & 83 $\pm$ 3 [a] \\
NGC4526	 & 	SAB(s)0	 & 617 $\pm$ 5 & 16.44 & 6.47 $\pm$ 0.02 & 6.46 $\pm$ 0.02 & -24.63 $\pm$ 0.02 & 213 $\pm$ 9 [a] \\
NGC4527	 & 	SAB(s)bc	 & 1736 $\pm$ 1 & 14.20 & 6.93 $\pm$ 0.02 & 6.95 $\pm$ 0.04 & -23.81 $\pm$ 0.04 & 135 $\pm$ 8 [a] \\
NGC4535	 & 	SAB(s)c	 & 1964 $\pm$ 1 & 15.60 & 7.38 $\pm$ 0.05 & 8.02 $\pm$ 0.06 & -23.82 $\pm$ 0.06 & 102 $\pm$ 10 [a] \\
NGC4559	 & 	SAB(rs)cd	 & 814 $\pm$ 1 & 8.91 & 7.58 $\pm$ 0.05 & 7.61 $\pm$ 0.06 & -22.14 $\pm$ 0.06 & 49 $\pm$ 9 [a] \\
NGC4569	 & 	SAB(rs)ab;Sy &  -235 $\pm$ 4 & 17.00 & 6.58 $\pm$ 0.03 & 6.59 $\pm$ 0.05 & -24.66 $\pm$ 0.05 & 136 $\pm$ 3 [a] \\
NGC4579	 & 	SAB(rs)b;Sy & 1517 $\pm$ 1 & 23.00 & 6.49 $\pm$ 0.03 & 6.38 $\pm$ 0.04 & -25.43 $\pm$ 0.04 & 165 $\pm$ 4 [a] \\
NGC4654	 & 	SAB(rs)cd	 & 1036 $\pm$ 1 & 13.10 & 7.74 $\pm$ 0.03 & 7.75 $\pm$ 0.06 & -22.83 $\pm$ 0.06 & 48 $\pm$ 9 [a] \\
NGC5005	 & 	SAB(rs)bc;Sy & 946 $\pm$ 5 & 14.60 & 6.44 $\pm$ 0.02 & 6.42 $\pm$ 0.04 & -24.40 $\pm$ 0.04 & 172 $\pm$ 8 [a] \\
M83		 & 	SAB(s)c	 & 513 $\pm$ 2 & 4.61 & 4.62 $\pm$ 0.02 & 4.57 $\pm$ 0.05 & -23.75 $\pm$ 0.05 & $\cdots$ \\
M101	 & 	SAB(rs)cd	 & 241 $\pm$ 2 & 7.00 & 5.51 $\pm$ 0.05 & 5.43 $\pm$ 0.05 & -23.80 $\pm$ 0.05 & 24 $\pm$ 9 [a] \\
NGC5746	 & 	SAB(rs)b	 & 1728 $\pm$ 2 & 34.70 & 6.88 $\pm$ 0.02 & 6.82 $\pm$ 0.04 & -25.89 $\pm$ 0.04 & 200 $\pm$ 8 [a] \\
NGC5985	 & 	SAB(r)b;Sy	 & 2522 $\pm$ 3 & 61.10 & 8.15 $\pm$ 0.04 & 8.07 $\pm$ 0.06 & -25.86 $\pm$ 0.06 & 158 $\pm$ 8 [a] \\
NGC6384	 & 	SAB(r)bc	 & 1665 $\pm$ 1 & 20.70 & 7.53 $\pm$ 0.04 & 7.42 $\pm$ 0.05 & -24.16 $\pm$ 0.05 & 124 $\pm$ 7 [a]\\
\multicolumn{8}{c}{------SB Galaxies------}\\
NGC613	 & 	SB(rs)bc	 & 1481 $\pm$ 5 & 15.40 & 7.03 $\pm$ 0.03 & 7.01 $\pm$ 0.04 & -23.92 $\pm$ 0.04 & 126 $\pm$ 19 [b] \\
NGC1097	 & 	SBb;Sy1	 & 1271 $\pm$ 3 & 24.90 & 6.25 $\pm$ 0.03 & 6.23 $\pm$ 0.03 & -25.75 $\pm$ 0.03 & 195 $\pm$	10 [b] \\
NGC1291	 & 	SB0/a		 & 839 $\pm$ 2 & 9.08 & 5.66 $\pm$ 0.02 & 5.37 $\pm$ 0.02 & -24.42 $\pm$ 0.02 & 162 $\pm$	18 [b] \\
NGC1365	 & 	SBb(s)b;Sy	 & 1636 $\pm$ 1 & 17.80 & 6.37 $\pm$ 0.04 & 6.40 $\pm$ 0.04 & -24.85 $\pm$ 0.04 & 151 $\pm$ 20 [b] \\
NGC1433	 & 	SB(rs)ab;Sy2 & 1076 $\pm$ 1 & 9.04 & 7.06 $\pm$ 0.04 & 7.01 $\pm$ 0.04 & -22.76 $\pm$ 0.04 & 113 $\pm$ 3 [b] \\
NGC1672	 & 	SB(r)bc;Sy2  & 1331 $\pm$ 3 & 11.40 & 7.02 $\pm$ 0.03 & 6.94 $\pm$ 0.03 & -23.35 $\pm$ 0.03 & 111 $\pm$ 3 [b] \\
NGC2903	 & 	SB(s)d		 & 550 $\pm$ 1 & 10.40 & 6.04 $\pm$ 0.02 & 5.96 $\pm$ 0.05 & -24.12 $\pm$ 0.05 & 89 $\pm$ 4 [a] \\
NGC3198	 & 	SB(rs)c	 & 660 $\pm$ 1 & 14.50 & 7.78 $\pm$ 0.04 & 7.74 $\pm$ 0.06 & -23.06 $\pm$ 0.06 & 46 $\pm$ 9 [a] \\
NGC3319	 & 	SB(rs)cd	 & 739 $\pm$ 1 & 13.40 & 10.07 $\pm$ 0.05 & 9.11 $\pm$ 0.06 & -21.52 $\pm$ 0.06 & 87 $\pm$ 9 [a] \\
NGC3351	 & 	SB(r)b;HII	 & 779 $\pm$ 1 & 9.30 & 6.66 $\pm$ 0.04 & 6.56 $\pm$ 0.04 & -23.28 $\pm$ 0.04 & 120 $\pm$ 9 [a] \\
M108	 & 	SB(s)cd	 & 699 $\pm$ 1 & 8.80 & 7.04 $\pm$ 0.02 & 7.11 $\pm$ 0.02 & -22.99 $\pm$ 0.02 & 79 $\pm$ 10 [a] \\
NGC3953	 & 	SB(r)bc	 & 1052 $\pm$ 1 & 15.40 & 7.05 $\pm$ 0.03 & 7.00 $\pm$ 0.05 & -23.94 $\pm$ 0.05 & 116 $\pm$ 3 [a] \\
NGC4442	 & 	SB(s)0		 & 547 $\pm$ 5 & 15.30 & 7.29 $\pm$ 0.02 & 7.28 $\pm$ 0.03 & -23.64 $\pm$ 0.03 & 187 $\pm$ 8 [a] \\
M91		 & 	SBb(rs);Sy	 & 486 $\pm$ 4 & 17.90 & 7.12 $\pm$ 0.03 & 6.76 $\pm$ 0.04 & -24.51 $\pm$ 0.04 & 113 $\pm$ 9 [a] \\
NGC4593	 & 	SB(rs)b;Sy1	 & 2492 $\pm$ 6 & 38.50 & 7.98 $\pm$ 0.03 & 7.78 $\pm$ 0.04 & -25.14 $\pm$ 0.04 & 105 $\pm$ 5 [b] \\
NGC4731	 & 	SB(s)cd	 & 1491 $\pm$ 1 & 12.40 & 9.78 $\pm$ 0.06 & 9.39 $\pm$ 0.06 & -21.08 $\pm$ 0.06 & $\cdots$ \\
NGC4754	 & 	SB(r)0-	 & 1351 $\pm$ 5 & 15.90 & 7.41 $\pm$ 0.03 & 7.34 $\pm$ 0.03 & -23.66 $\pm$ 0.03 & 185 $\pm$ 4 [a] \\
NGC4945	 & 	SB(s)cd;Sy2 & 563 $\pm$ 3 & 3.58 & 4.48 $\pm$ 0.02 & 4.44 $\pm$ 0.05 & -23.33 $\pm$ 0.05 & 134	$\pm$ 20 [b] \\
M51b	 & 	SB0;pec	 & 465 $\pm$ 1 & 7.31 & 6.25 $\pm$ 0.03 & 6.34 $\pm$ 0.04 & -22.98 $\pm$ 0.04 & 125 $\pm$ 8 [a] \\
NGC5792	 & 	SB(rs)b	 & 1926 $\pm$ 1 & 26.20 & 7.71 $\pm$ 0.03 & 7.66 $\pm$ 0.04 & -24.44 $\pm$ 0.04 & $\cdots$ \\
NGC5850 & 	SB(r)b	 & 2545 $\pm$ 1 & 17.80 & 8.10 $\pm$ 0.04 & 7.89 $\pm$ 0.04 & -23.36 $\pm$ 0.04 & 140 $\pm$ 7 [a] \\
NGC7582 &	SB(s)ab;Sy2 & 1575 $\pm$ 7 & 22.30 & 7.32 $\pm$ 0.02 & 7.36 $\pm$ 0.03 & -24.38 $\pm$ 0.03 & 113 $\pm$ 3 [b] \\
\hline
\multicolumn{8}{l}{}\\
\multicolumn{8}{l}{\textit{Table 1 continued}. \textbf{Notes.} (1) Galaxy name. (2) Hubble type from \citet{Jarrett_et_al_2003}. (3) Heliocentric}\\
\multicolumn{8}{l}{ velocity from NED. (4) Mean of distance determinations from NED (based on primary distance indicators if }\\
\multicolumn{8}{l}{available; see also Paper II for more references). (5) Total magnitude in $K_{s}$ band from 2MASS. (6) Total mag-}\\
\multicolumn{8}{l}{nitude in $K_s$ band measured in this work. (7) Total absolute magnitude in $K_s$ band measured in this work.}\\
\multicolumn{8}{l}{Magnitudes for ETGs measured in this work were derived using 2MASS image tiles (see \S \ref{LGA_VS_tile}). Uncertainties}\\
\multicolumn{8}{l}{in our magnitudes are solely from GALFIT models. (8) Stellar velocity dispersion data from the next}\\
\multicolumn{8}{l}{ references: (a) \citet{Ho_et_al_2009}; (b) Hyperleda database \citep{Paturel_et_al_2003}.}\\
\end{tabular}
\end{table*}

\section{Methodology}\label{Met}

\subsection{2D Decomposition with GALFIT}

We used the GALFIT\footnote{Software available at \url{https://users.obs.carnegiescience.edu/peng/work/galfit/galfit.html}} \citep{Peng_et_al_2002, Peng_et_al_2010} algorithm to extract the structural parameters of the galaxies, such as total magnitudes, effective radii, concentration indices, among others, from the surface brightness distribution modelling. GALFIT performs a photometric two-dimensional decomposition on digital images of galaxies, stars or other astronomical sources. The modelling is carried out through parametric functions, such as S\'ersic, Exponential, Nuker, Gaussian, Moffat, Ferrer, among others, and it can also be used to perform multiple fits on several objects simultaneously. Some wrapper scripts have been implemented to fit galaxies in crowded regions, such as in clusters of galaxies \citep[e.g.,][]{Anorve_PhD_Thesis,2012MNRAS.422..449B}.

GALFIT uses a least-squares Levenberg-Marquardt minimization, while the goodness of the fit is computed by the chi-square, $\chi^{2}$ statistics, then calculates the parameters for the next step and continues iterating until the $\chi^{2}$ is minimised. The indicator of the goodness of the fit is the reduced $\chi^{2}$, $\chi^{2}_{\nu}$:

\begin{equation}
\chi^{2}_{\nu} = \frac{1}{N_{dof}}\sum_{x=1}^{nx}\sum_{y=1}^{ny}\frac{\left[f_{img}(x,y) - f_{mod}(x,y)\right]^{2}}{\sigma(x,y)^{2}},
\label{chi2}
\end{equation}
where $N_{dof}$ is the number of degrees of freedom in the fit; $nx$ and $ny$ are the image dimensions; $f_{img}(x,y)$ is the value of the $(x,y)$ pixel of the galaxy image; $f_{mod}(x,y)$ is the value of the corresponding pixel of the PSF-convolved model image generated at each iteration; and $\sigma(x,y)$ is the ``sigma'' image, which is the error on each pixel and is generated internally by GALFIT.

GALFIT reads the image header to access the exposure time and $GAIN$, while the magnitude zero-point is taken from the input parameters file. The image ADUs are converted into electrons using the $GAIN$ parameter, so that $ADU$ $\times$ $GAIN=electrons$. The magnitude zero-point is used to convert pixel counts and fluxes into a physical magnitude. The PSF is provided by the user. A good PSF is generated by using bright, isolated non-saturated stars. GALFIT convolves the PSF image with the model during the fit to account for the effects of seeing and/or also to fit a PSF as a model component. The PSF image is obtained using the SExtractor \citep{Bertin-Arnouts_1996} ``objects'' image to select the star among the potential candidates. We also have excluded bad pixels or objects from the fit by manually masking them (in most of the images of our sample the bright stars were already masked), using the SExtractor ``segmentation'' image, which is modified when such pixels or objects are selected in order to set the all non-zero valued pixels that will be ignored during the fit \citep[see][for further details]{Peng_et_al_2010}. Parameters are allowed to vary during the fit.

Initial parameters are generated by SExtractor and then are parsed to GALFIT, namely: the centroid $x$ and $y$ pixel positions, initial total magnitude, initial axial ratio and effective radius. The S\'ersic index is initially set to $n$ $\sim$ 2, depending on the morphological type of the galaxy.

\subsection{Model Selection}

Based on the morphological type reported by 2MASS \citep{Jarrett_et_al_2003} and listed in column 4 of Table \ref{Table_sample}, we considered mainly three models when adjusting the surface brightness of galaxies: usually a single component represented by a S\'ersic profile for elliptical galaxies; bulge and disc components of lenticular and spiral galaxies are modelled by S\'ersic and exponential functions, respectively; while bulge, disc and bar components are considered for barred galaxies, where  bars are modelled also by a S\'ersic profile. For those sources classified as AGN, their models can include an additional PSF component to fit the unresolved nuclear component associated with the AGN contribution. A comprehensive discussion on the applicability of different models is given by \citet{Graham_2013}.

GALFIT parameterises the surface brightness distribution of galaxies and compact sources using axially symmetric profiles, whose radial distribution is expressed by generalised ellipses \citep{1990MNRAS.245..130A}:
\begin{equation}
r=\left(\lvert x \rvert^{(c+2)} + \left\lvert \frac{y}{q} \right\rvert^{(c+2)} \right)^{\frac{1}{(c+2)}},
\end{equation}
where $q$ is the ratio of the major axis to the minor axis, while $c$ indicates the boxiness/diskiness of the ellipses. 

The popularity of the S\'ersic profile \citep{1963BAAA....6...41S, Sersic_1968} has grown since its revival during the 90's \citep[e.g.,][]{1993MNRAS.265.1013C}. It has been implemented in GALFIT by the following expression:

\begin{equation}
I(r) = I_{e}\exp\left[-\kappa\left(\left(\frac{r}{r_{e}}\right)^{1/n} - 1\right)\right],
\label{Ser_profile}
\end{equation}

where $I(r)$ is the surface brightness at the radius $r$ and $\kappa$ is a parameter coupled to the S\'ersic index $n$, a measure of the concentration of the light profile, in such way that $I_{e}$ is the surface brightness at the effective radius $r_{e}$ (radius where half of the total flux is within it). When $n$ = 4, then $\kappa=7.66925$ and we have the \citet{deVaucouleurs_1948} profile. 

The exponential profile \citep{Freeman_1970} is also a special case of the S\'ersic function when $n$ = 1. It is given by the following expression:
\begin{equation}
I(r) = I_{0}\exp\left(-\frac{r}{r_{s}}\right),
\label{exp_disc_profile}
\end{equation}
where $I_{0}$ is the central surface brightness and $r_{s}$ is the scale length of the disc. $r_s$ is related to the effective radius $r_{e}$ by $r_{e}$ = 1.678$r_{s}$.

To model the bar component, as mentioned above, we used a S\'ersic profile with $n$ $\leq$ 0.5, which is often used in literature for this purpose \citep[e.g.,][]{Greene_et_al_2008, Peng_et_al_2010}.

Fits to the surface brightness distribution of galaxies were done following the steps described below:

\begin{enumerate}
    \item SExtractor was used to generate input parameters. Then, we first modelled with  S\'ersic profile, regardless of the galaxy morphology.
    
    \item Elliptical galaxies are expected to be well described by a S\'ersic profile, but we do not discard fitting an additional component (double S\'ersic or exponential profile), to respond to the possible existence of substructures that resembles extended stellar envelopes or embedded discs \citep[e.g.,][]{lasker_et_al_2014}. Hence, some ellipticals in our sample have been fitted with an additional component to the S\'ersic component described above and, when it is the case, we also call bulge to that central component having a higher $n$ (also a checking on $q$ ratio and morphology of the additional substructure is done).
    
    \item For disc galaxies, lenticulars and normal spirals, an additional run is made with a single exponential disc component. 
    
    \item For barred spiral galaxies is proceeded in the same way, adding a third component to fit the bar. In the case of intermediate spirals (SAB), some of them are well fitted with two components, others may need three components, while in some cases the models between a SAB galaxy with two and three components are quite similar. 
    
    \item The sky background component is also considered as a free parameter in the GALFIT runs.

    \item We include all the components considered previously (for instance, S\'ersic, disc and/or bar models depending on the morphological type of the galaxy), using their outputs as initial parameters,  to run  a simultaneous fit to arrive at the final model. 

    \item The selection of the best-fitting model is based on the goodness of the fit (reduced-$\chi^2$) and through a visual inspection of residual image. We improved our models by including additional components to account for symmetrical structures identified in the residual images (generally bars or an extra component for some ellipticals).
\end{enumerate}

\subsection{Uncertainties}
\label{Uncertainties}

The uncertainties of parameters returned by GALFIT underestimate the true error bars, which are typically $\sim$0.01 mag and $\sim$0.05 arcsec for $r_{e}$ \citep[e.g.,][]{Haussler_et_al_2007}. Therefore, we have used  a different approach to derive more realistic uncertainties. We refer the reader to the Appendix A for a detailed description and discussion on the procedure for estimating the final uncertainties, reported in this paper.

\subsection{Surface Brightness Profiles}
\label{SB_profiles}
We used EllipSect\footnote{Software available at  \url{http://github.com/canorve/EllipSect}} \citep{Anorve_2020} to generate the surface brightness profiles of galaxies from GALFIT's outputs. Besides, EllipSect is able to extract and compute the absolute magnitude, luminosity, flux, total apparent magnitude, Akaike Information Criterion (AIC) and Bayesian Information Criterion (BIC), among other parameters. 1D profiles are useful to evaluate the choice of the final model generated through the 2D decomposition, since the need of additional components may be noticed when comparing the surface brightness profiles of the galaxy and its model.

Examples of the photometric decomposition carried out with GALFIT are shown in Figure \ref{varias_3bands} along with the 1D profiles generated using EllipSect for the galaxies NGC4125, M51a and M91.

\subsection{Extinction and k-Correction}

The NIR magnitudes  were corrected for Galactic extinction according to \citet{Schlegel_et_al_1998}, as provided by the NASA/IPAC Infrared Science Archive\footnote{\url{https://irsa.ipac.caltech.edu/applications/DUST/}}.

Additionally, we applied $k$-corrections to the magnitudes using the galaxy's redshift and the $J$-$K_s$ and $H$-$K_s$ colours, according to the prescription from \citet{Chilingarian_et_al_2010}, which is also available online\footnote{\url{http://kcor.sai.msu.ru}}. \citet{Chilingarian_et_al_2010} scheme to generate $k$-correction is comparable to the one developed by \citet{2001MNRAS.326..745M}.

\begin{figure*}
\begin{center}
\includegraphics[width=17.65 cm]{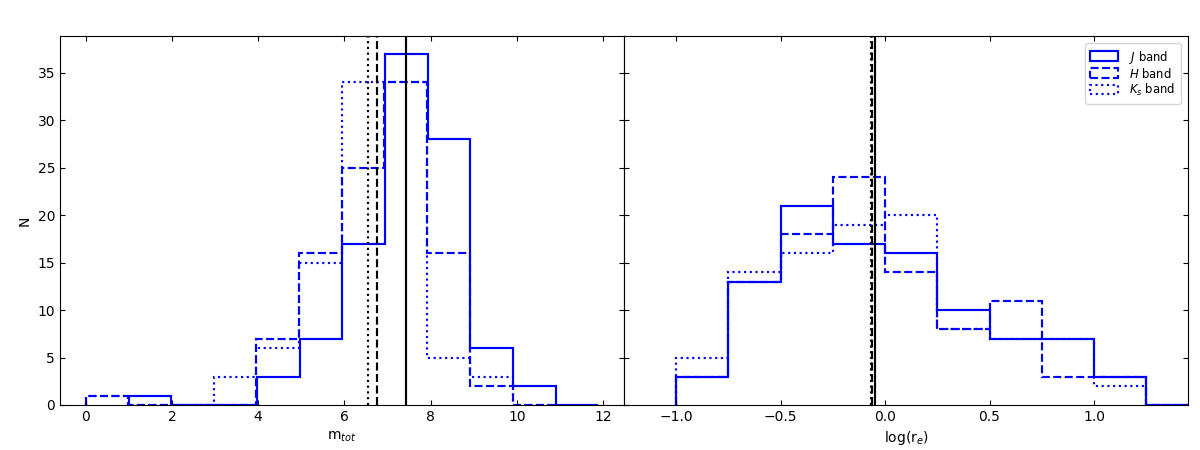}
\caption[m_tot_r_eff-histos]{\footnotesize Distribution magnitudes (left panel) and effective radii obtained from our photometric analysis. Solid line is for $J$ band, while dashed and dotted lines are for $H$ and $K_s$ bands, respectively. In the same way, vertical lines indicate the mean of each distribution.}
\label{m_tot_r_eff-histos}
\end{center}
\end{figure*}

\begin{figure*}
\begin{center}
\includegraphics[width=17.5 cm]{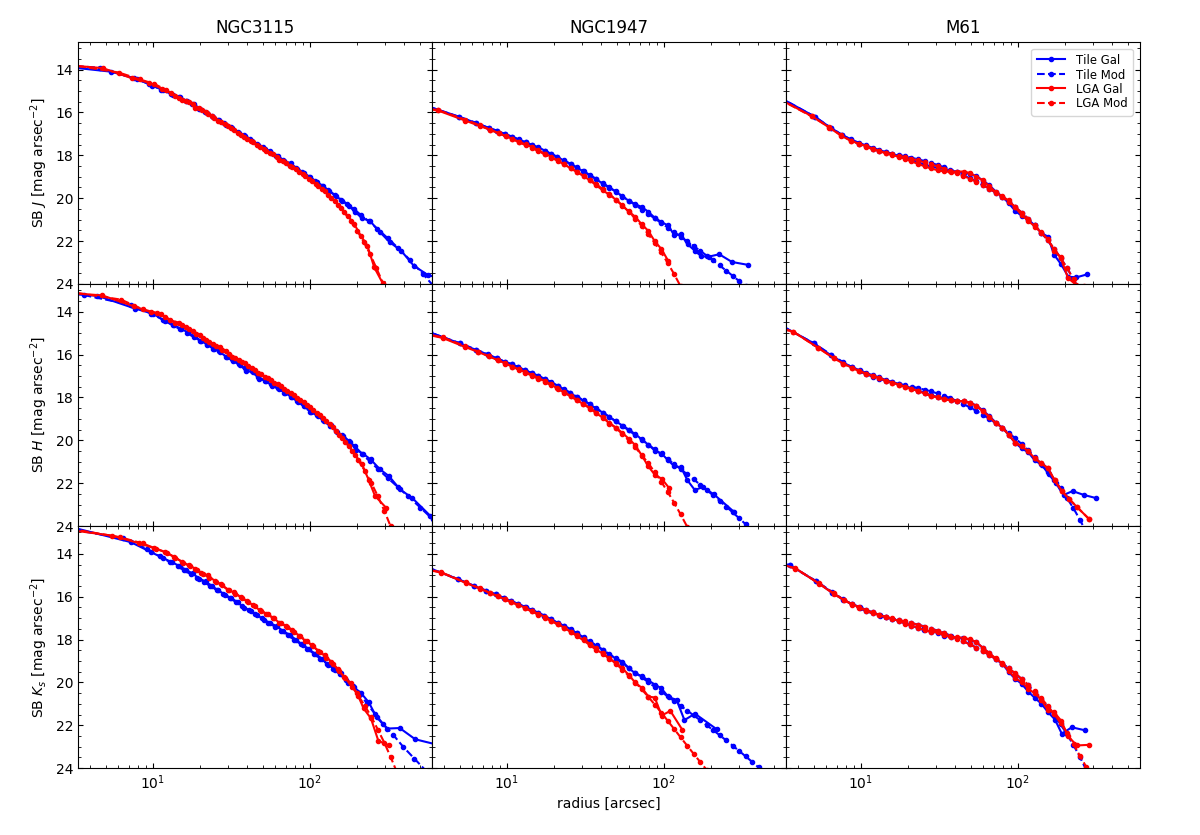}
\caption[plot_SB_gals]{\footnotesize Surface brightness profiles using LGA mosaics (red points and lines) and IRSA image tiles (blue points and lines) for 2MASS data, for the ETGs NGC3115 (left) and  NGC1947 (center), and the LTG M61 (right). Solid lines represent the galaxy profile, while dashed lines are for the models generated with GALFIT. From top to bottom panel are shown $J$, $H$ and $K_s$ bands, respectively. An abrupt drop down is seen in the profiles for ETGs using LGA mosaics, while for the LTG the profiles are consistent using either LGA mosaics or 2MASS IRSA image tiles.}
\label{plot_SB_gals}
\end{center}
\end{figure*}

\begin{figure*}
\begin{center}
\includegraphics[width=17.65 cm]{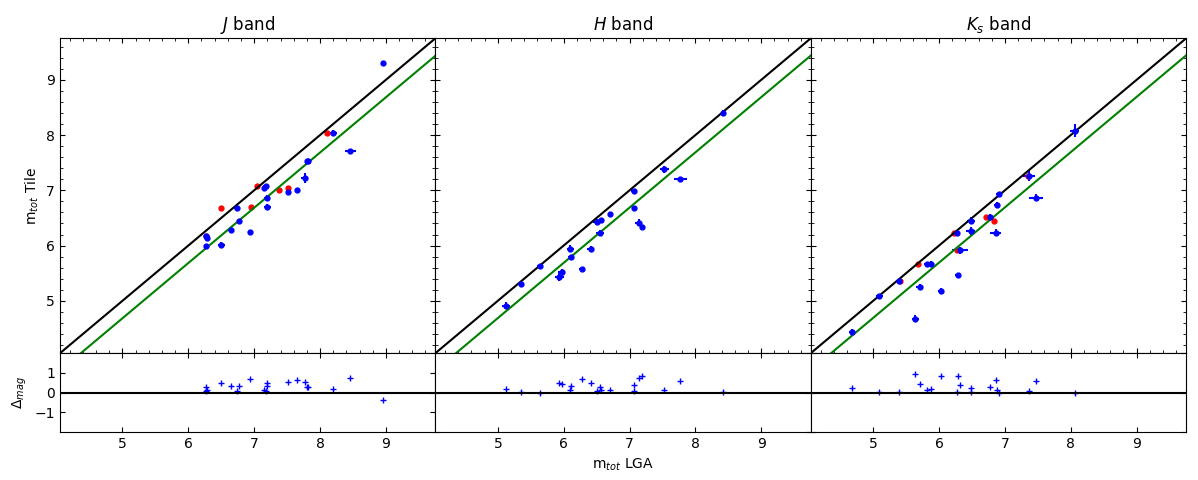}
\caption[1-1_mag_all]{\footnotesize Comparison between total magnitudes for ETGs in our sample using 2MASS data for LGA mosaics and image tiles in $J$, $H$ and $K_s$ bands. Black lines represent the one-to-one correspondence, while the green ones represent an offset of $\sim$ 0.35 mag per band. Red points are galaxies in common with \citet{Schombert_2011}}
\label{1-1_mag_all}
\end{center}
\end{figure*}

\begin{figure*}
\begin{center}
\includegraphics[width=17.65 cm]{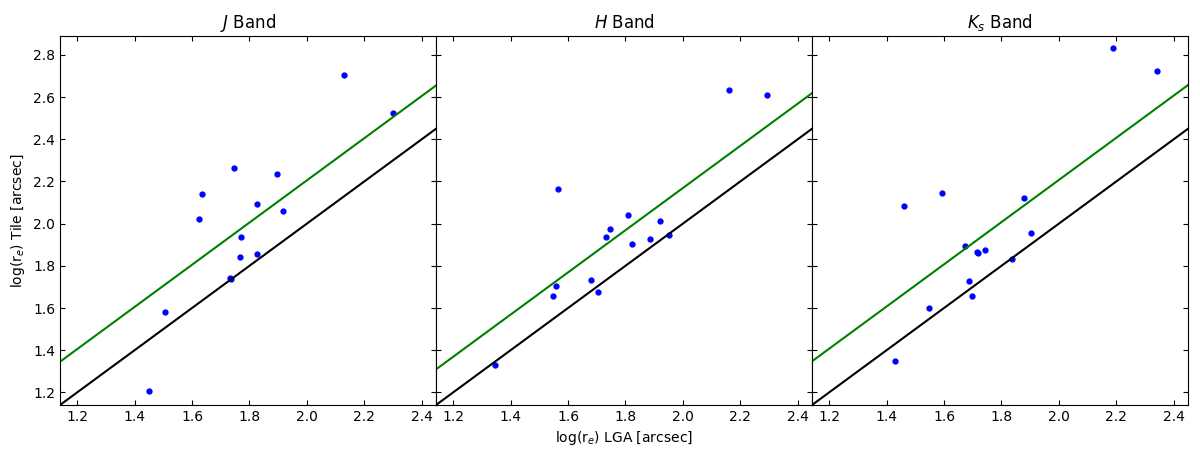}
\caption[1-1_reff]{\footnotesize Same as previous image, but for effective radii of galaxies estimated using LGA mosaics and IRSA image tiles for E galaxies in our sample. Black lines represent the one-to-one correspondence, while the green ones represent an offset of $\sim$ 0.20 dex per band.}
\label{1-1_reff}
\end{center}
\end{figure*}

\begin{figure}
\includegraphics[width=8.25 cm]{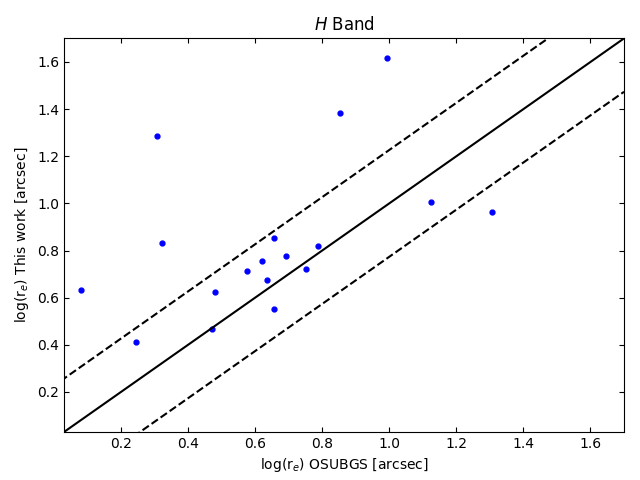}
\caption[1-1_reff_OSU]{\footnotesize Comparisons between effective radii of bulges estimated in this work with those ones from \citet{Laurikainen_et_al_2004_2} using data from OSUBSGS in $H$ band. Dashed lines represent the 1$\sigma$ scatter.}
\label{1-1_reff_OSU}
\end{figure}

\begin{figure}
\includegraphics[width=8.25 cm]{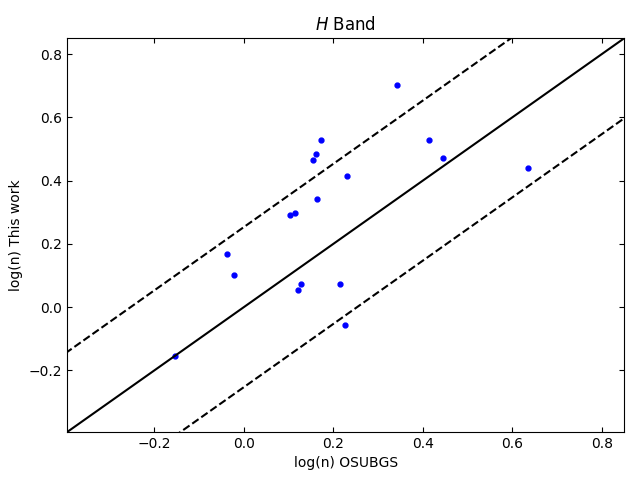}
\caption[1-1_n_OSU]{\footnotesize Same as previous image, but for Sersic index of bulges estimated in this work and those ones from \citet{Laurikainen_et_al_2004_2}.}
\label{1-1_n_OSU}
\end{figure}

\begin{figure*}
\begin{center}
\includegraphics[width=17.6 cm]{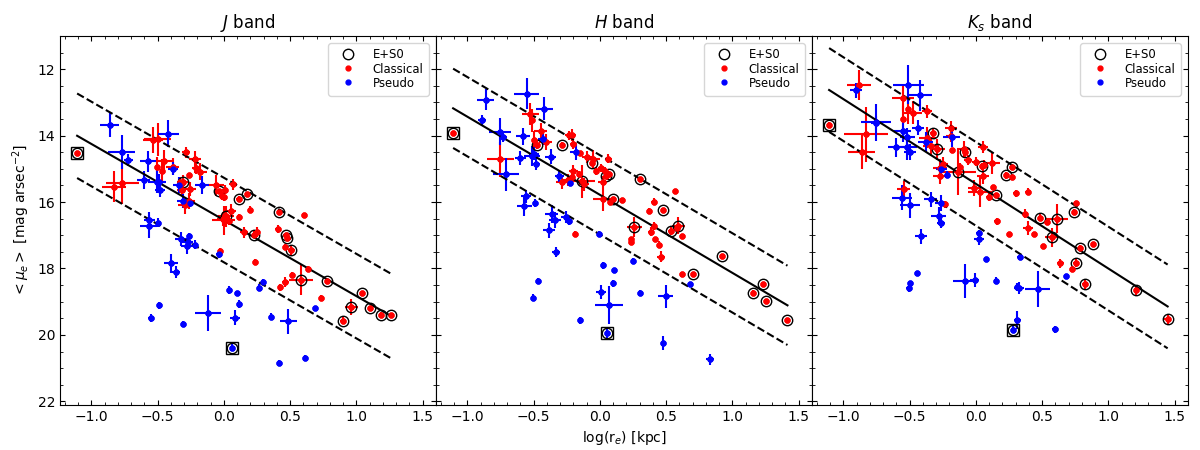}
\caption[KR_final]{\footnotesize Kormendy Relation (KR) for E+S0 galaxies and bulges in our sample. Red filled circles represent classical bulges, while the blue ones are pseudobulges according to our final classification. Black dashed lines represent 3$\sigma$ to the fit of E+S0 galaxies in our sample. The three bands of 2MASS are shown. Objects in black circle represent the E+S0 galaxies in our sample. M32 and M110 are highlighted in black squares (see \S\ref{Ind_cases}).}
\label{KR_final}
\end{center}
\end{figure*}

\begin{figure*}
\begin{center}
\includegraphics[width=17.65 cm]{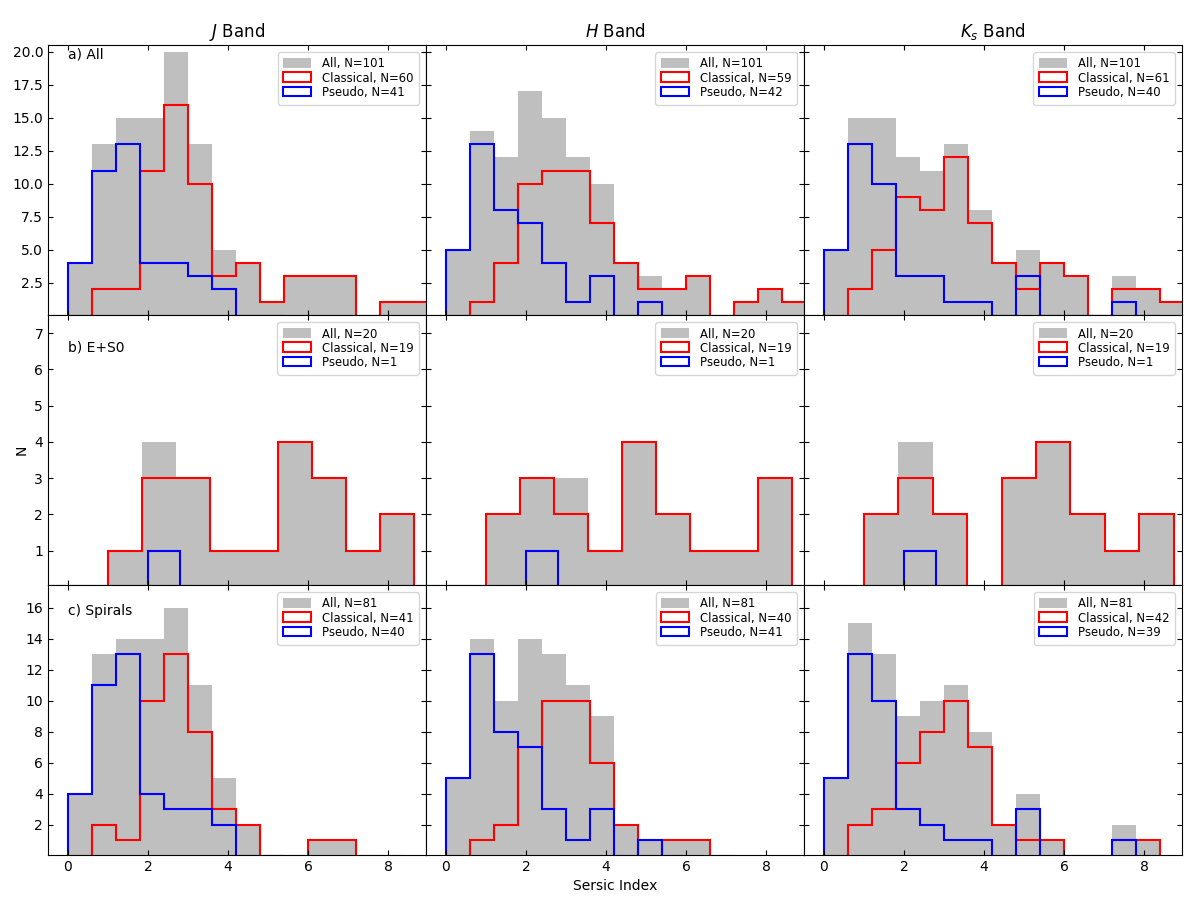}
\caption[n_subplots_Galaxies]{\footnotesize Distribution of the S\'ersic index according to the (sub)sample of galaxies considered. Panels are from top to bottom: a) All sample, b) E+S0 galaxies and c) spiral galaxies (S, SAB and SB types), respectively. From left to right: $J$, $H$ and $K_s$ bands, respectively. In all cases gray histograms indicate the total (sub)sample, while the red and blue ones are for classical and pseudo bulges, respectively.}
\label{n_subplots_Galaxies}
\end{center}
\end{figure*}

\begin{figure*}
\begin{center}
\includegraphics[width=17.65 cm]{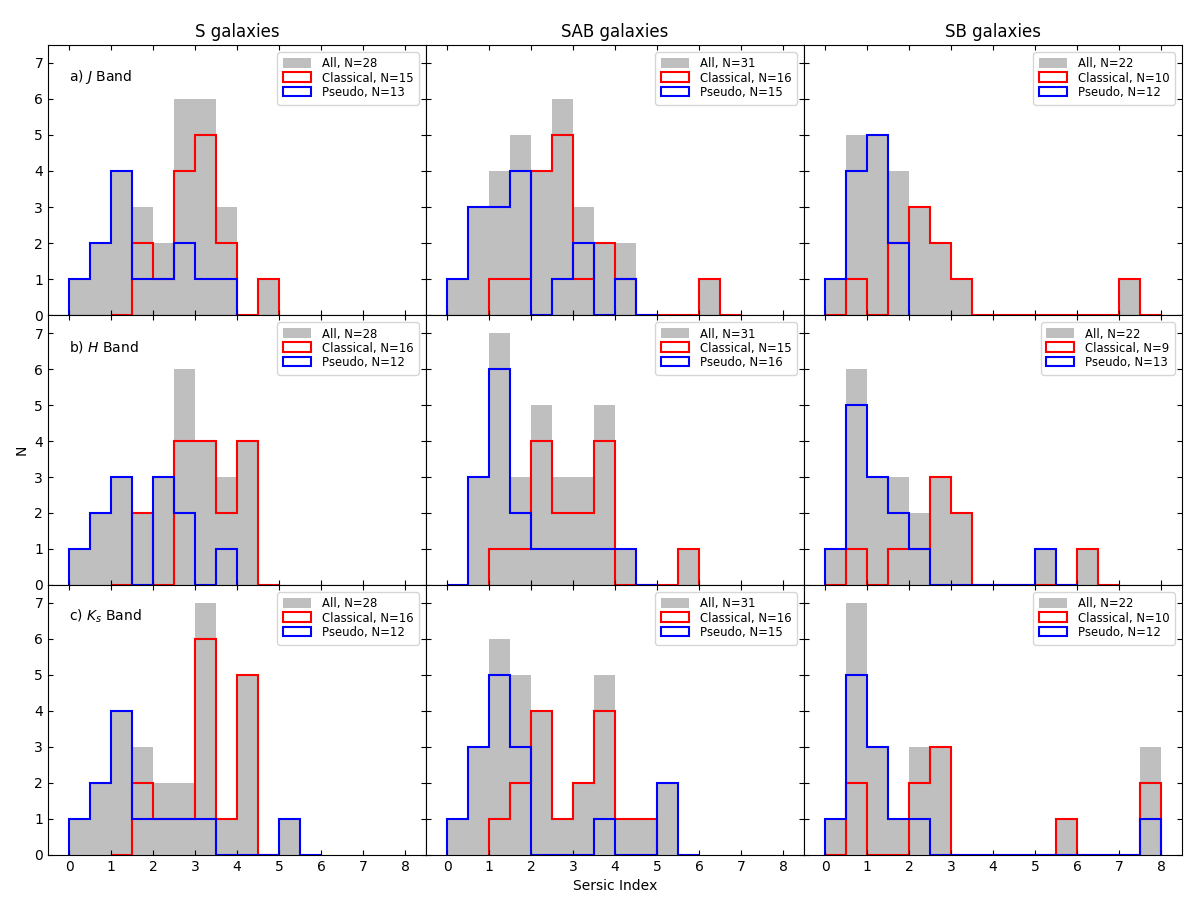}
\caption[n_subplots_Spirals]{\footnotesize Distribution of the S\'ersic index for spiral galaxies. Panels are from top to bottom: a) $J$, b) $H$ and c) $K_s$ bands, respectively. From left to right: S, SAB and SB Hubble types, respectively. Colours for histograms are the same as in Figure \ref{n_subplots_Galaxies}.}
\label{n_subplots_Spirals}
\end{center}
\end{figure*}

\begin{figure*}
\begin{center}
\includegraphics[width=17.65 cm]{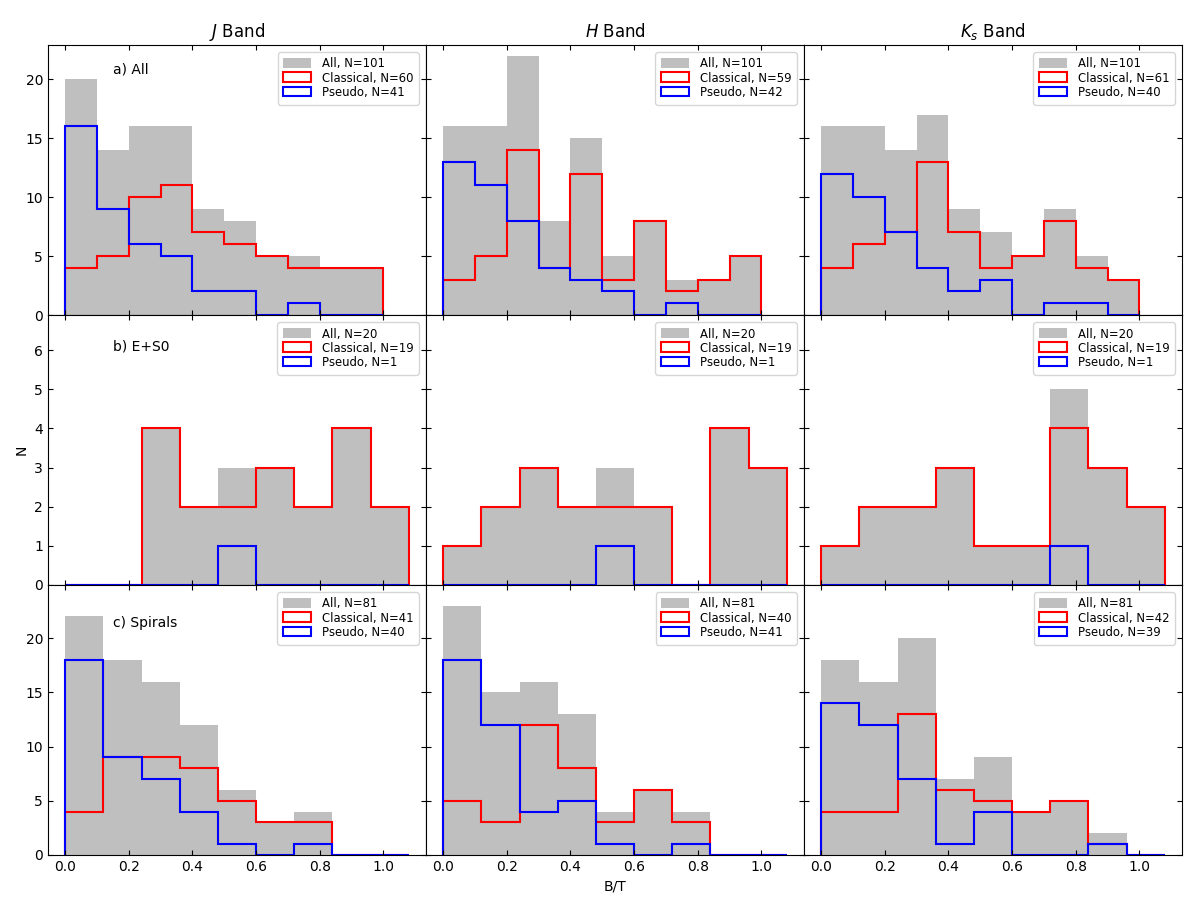}
\caption[BT_subplots_Galaxies]{\footnotesize Distribution of the $B/T$ ratio according to the (sub)sample of galaxies considered. Panels are from top to bottom: a) All sample, b) E+S0 galaxies and c) spiral galaxies (S, SAB and SB types), respectively. From left to right: $J$, $H$ and $K_s$ bands, respectively. Colours for histograms are the same as in Figure \ref{n_subplots_Galaxies}.}
\label{BT_subplots_Galaxies}
\end{center}
\end{figure*}

\begin{figure*}
\begin{center}
\includegraphics[width=17.65 cm]{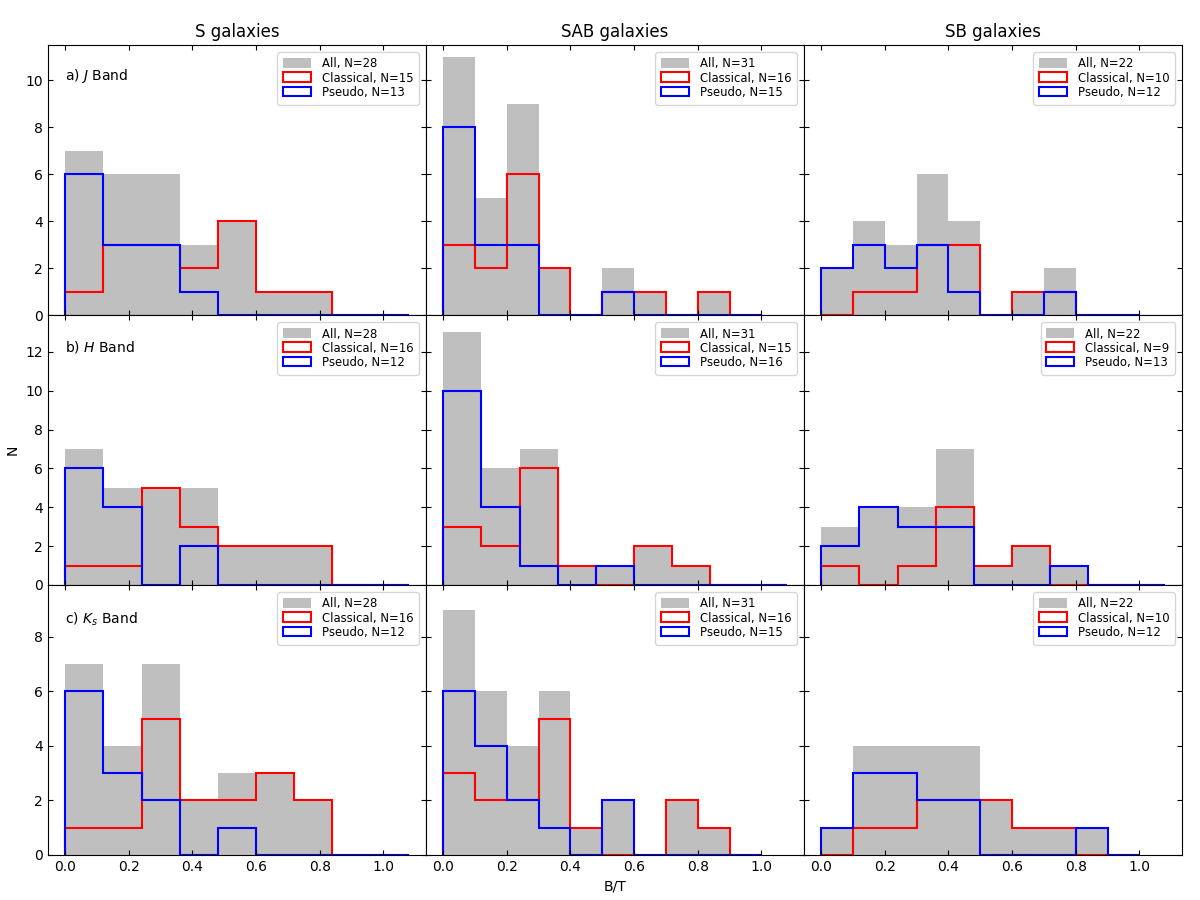}
\caption[BT_subplots_Spirals]{\footnotesize Distribution of the $B/T$ ratio for spiral galaxies. Panels are from top to bottom: a) $J$, b) $H$ and c) $K_s$ bands, respectively. From left to right: S, SAB and SB Hubble types, respectively. Colours for histograms are the same as in Figure \ref{n_subplots_Galaxies}.}
\label{BT_subplots_Spirals}
\end{center}
\end{figure*}

\begin{figure*}
\begin{center}
\includegraphics[width=17.65 cm]{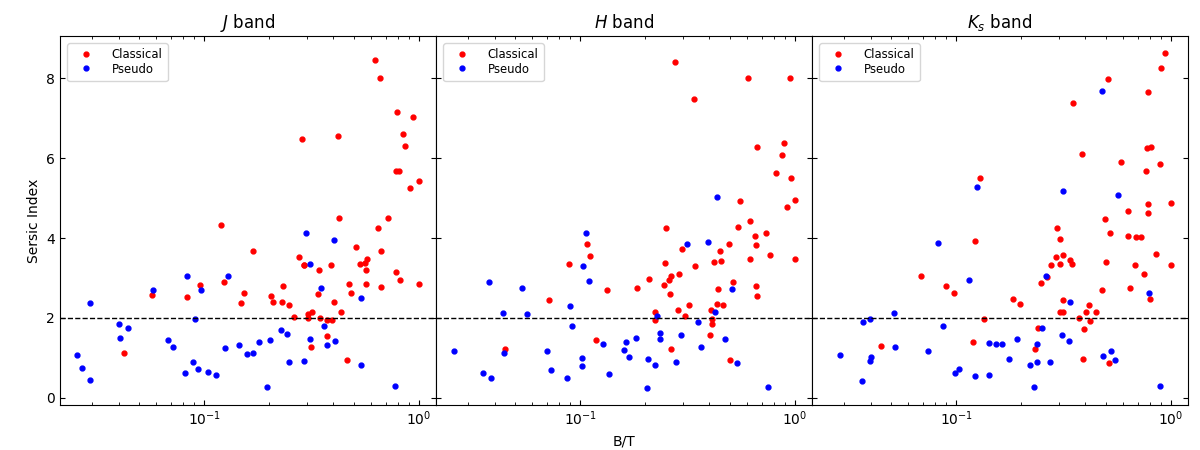}
\caption[n-BT]{\footnotesize Relation between $B/T$ and S\'ersic index for the galaxies in our sample. According to our classification for bulges, red points represent classical bulges, while the blue ones are pseudobulges. Horizontal dashed lines indicate $n$=2.}
\label{n-BT}
\end{center}
\end{figure*}

\begin{figure*}
\begin{center}
\includegraphics[width=17.65 cm]{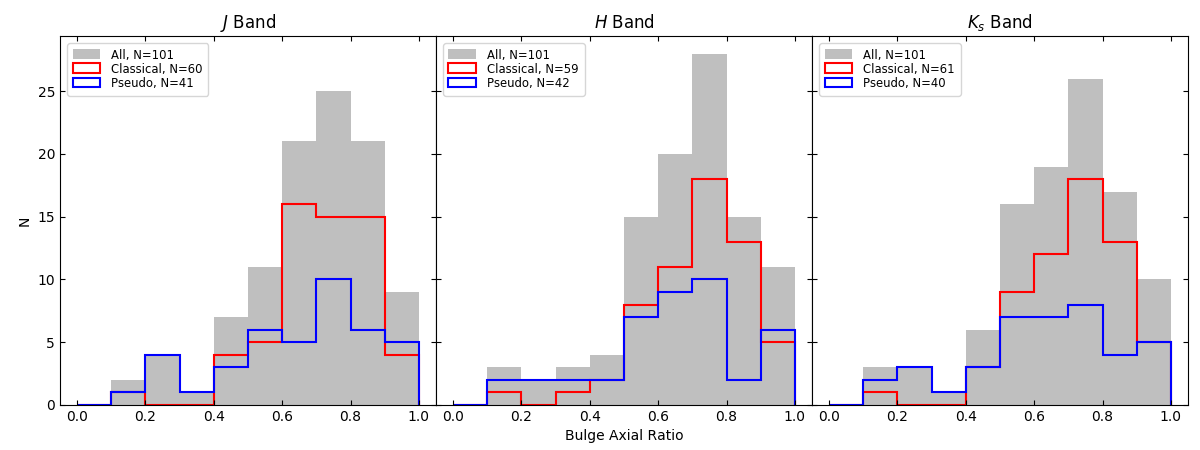}
\caption[q_bulges]{\footnotesize Distribution of the axial ratio $q$ for the galaxies in our sample.}
\label{q_bulges}
\end{center}
\end{figure*}

\subsection{Issues with the Large Galaxy Atlas Photometry}\label{LGA_VS_tile}

We have performed a comparison between data from the 2MASS by using data from the LGA mosaics and IRSA image tiles. We found disagreement among the structural parameters and the integrated total magnitudes for ETGs. The discrepancy was produced by the sky subtraction during the LGA mosaics generation. We show in Figure \ref{plot_SB_gals}, the surface brightness profiles of the ETGs NGC3115 and NGC1947 and the LTG M61. It can be seen how the surface brightness profiles of ETGs, generated on LGA mosaics (red line) show a steep drop-down than the surface brightness generated on IRSA image tiles (blue line); while for the LTG M61, the results are indistinguishable either using the LGA mosaic or the image tile. The LGA mosaics were made available to us by Jarret (2012, private communication) in a manageable set, where bright objects around the target galaxy were masked. We initially worked with the 101 galaxies in our sample using the LGA mosaics.

We repeated our analysis for all the 20 ETGs (E+S0) in our sample using the IRSA image tiles, where bright objects were masked. Figure \ref{1-1_mag_all} shows the comparison between the total magnitudes of ETGs using both data sets. The total magnitudes measured on IRSA tiles are brighter, according the following offsets: 

\begin{center}
\begin{tabular}{ l  c  } 
Band & Offset\\
\hline
$J$  & $0.32 \pm 0.08$ mag\\ 
$H$  & $0.34 \pm 0.06$ mag \\
$K_s$  & $0.34 \pm 0.07$ mag\\ 
\hline 
\end{tabular}
\end{center}

These results are consistent with the $\sim$0.33 mag offset reported by  \citet{Schombert_2011} and \citet{Schombert_2012}, and  the mean 0.34 mag offset found in  the much deeper photometric $K_s$ study by \citet{lasker_et_al_2014}.  Besides, for the few galaxies that we have in common with  \citet{Schombert_2012}, our measurements for their reported $J$ and $K_s$  magnitudes (red points in Figure \ref{1-1_mag_all}) are in agreement. As expected, an offset of 0.20 dex in the effective radius ($r_e$) of the galaxy was found for elliptical galaxies between data from the LGA mosaics and IRSA image tiles. These offsets are shown in Figure \ref{1-1_reff} for the $J,H, K_s$ bands. Our results are consistent with \citet{Cappellari_et_al_2011}, who found a larger offset.

We should note that \citeauthor{Schombert_2012} generated image cutouts from the 2MASS survey images using their own scripts. Besides, the sky determinations were generated by a slightly different approach to ours, and the surface brightness measurements were done using a 1D technique. Nevertheless, despite all those differences, our reported total magnitudes and offsets are in close agreement with \citet{Schombert_2012}.

A correction for ETGs photometry using LGA mosaics can be generated from the offsets found above, at the risk of sacrificing precision. Therefore, for the rest of the paper we report the photometric analysis for the 20 ETGs in our sample, generated on images from the IRSA tile server. On the other hand, since the results for LTGs photometry using LGA mosaics are indistinguishable from those derived using IRSA tiles, hereafter, we report the photometric analysis for the 81 LTGs in our sample generated from LGA mosaics. 

As a conclusion, we recommend that when dealing with bright ETGs, to avoid using 2MASS LGA mosaics or the reported measurements based on them. For example, some of the $K_s$ total magnitudes adopted by the MASSIVE Survey \citep[][their Table 3]{2014ApJ...795..158M} might be underestimated, as indicated in this paper.

\section{Results}\label{Res}

\subsection{Photometry and Structural Properties}

In this section we report the results of our two-dimensional surface brightness modelling with GALFIT, including the structural parameters such as magnitudes for each component (Table 3), as well as effective radii, S\'ersic indices ($n$), ellipticities and position angles (Tables 4, 5 and 6 for each band of 2MASS). These tables (from 3 to 6) are presented only in digital format. In such tables, subscripts 1, 2 and 3 for parameters refer to bulge component (or the single component in the case of some E galaxies), disc or envelope (see below \S \ref{E_gals_double}) and bar, respectively. The $r_s$ value reported in digital tables for E galaxies with an additional S\'ersic component refers to their $r_e$.

As it can be seen below in Figure \ref{m_tot_r_eff-histos}, our results look quite similar in the three 2MASS bands, where the distributions of total magnitudes and effective radii are displayed. Thereby, this allows us to have a better control of the structural parameters modelled with GALFIT, since quite similar conclusions are reached regardless of the NIR band used.

\subsubsection{Two-Component Fits to Elliptical Galaxies} \label{E_gals_double}

Some elliptical galaxies were modelled using two S\'ersic components \citep[e.g.,][for a view on early attempts]{1989woga.conf..208C}. Thus, when talking about bulge parameters for the complete sample, we refer to the central component that has the highest S\'ersic index. 

In order to carry out a more robust selection of the final model for these E galaxies, we applied the Akaike Information Criterion \citep[AIC,][]{Akaike_1974} and Bayesian Information Criterion \citep[BIC,][]{Schwarz_1978} implemented in EllipSect. AIC and BIC are based on the $\chi^2$ statistics and the numbers of parameters used in the model to penalise the overfitting of parameters added to the model. Then, the model with the lowest AIC and BIC values is the selected.

Below, we show the $n$ values for the E galaxies with two components in $J$ band (also in the $H$ and $K_s$ bands is performed the same methodology for such E galaxies) using IRSA image tiles (see \S \ref{LGA_VS_tile}); full results are disclosed in Tables 3-6. Therefore, M32 (cE2) was modelled with a bulge component with $n$=4.25 plus a disc component. M87 (E+0-1) has $n$=1.27 for the central component, while the additional component, the extended stellar envelope, has $n$=0.99. The surface brightness model of M60 (E2) has a bulge component with $n$=2.09, while the extra component is also a S\'ersic with $n$=1.47. M105 (E1) was best fitted with a S\'ersic component with $n$=3.21 plus a disc exponential profile as an additional component. The surface brightness distribution of M110 (E5) was modelled with a double S\'ersic component, with $n$=2.51 and 0.97 for the bulge and the extended envelope, respectively. In the case of NGC1549 (E0-1), a S\'ersic component with $n$=6.31 was fitted along with a disc exponential profile. The model of NGC3377 (E5-6) has two S\'ersic components, with $n$ values of 6.55 and 2.69 for the bulge and extended envelope, respectively. NGC4125 (E6) was fitted with S\'ersic component with $n$=7.03, plus an exponential disc profile representing an embedded disc. The central component of NGC4697 (E6) has $n$=4.51, while the envelope has $n$=3.01. Finally, the surface brightness profile of NGC4976 (E4) was fitted with a double S\'ersic with $n$=8.47 and $n$=1.81 for the bulge and stellar envelope, respectively.

A recent study by \citet{2013ApJ...766...47H} has suggested the presence of three-components in the surface brightness modelling of ETGs in the optical band $V$: inner, intermediate and extended, respectively. Although the galaxies in our sample are close enough ($z\leq 0.01$) to allow enough angular resolution to recover the inner components introduced by \citet{2013ApJ...766...47H}, we have failed to recover those minor inner  components ($r_e \sim 1\; {\rm kpc}$), which account for just  0.1 to 0.15 of the total luminosity of the galaxy. Those inner and intermediate components might be related to the continuous star formation seen in recently quenched and blue-star forming E galaxies \citep{2020A&A...644A.117L}. Therefore,   because the NIR bands used in this study trace older stellar populations, preferentially, we are unable to recover the supposedly younger inner components. More detailed comparisons between optical and NIR observations  are needed to  test  the IFU analysis  of \citet{2020A&A...644A.117L} or the  scenario proposed by \citet {2013ApJ...768L..28H}.

\subsubsection{\textbf{Comparison with previous works}}\label{Comparison_data}

Similarly to the comparisons of magnitudes presented above with \citet{Schombert_2012}, we also compared with the results from \cite{deJong_1994} and \cite{Tully_et_al_1996} for spiral galaxies in common observed in the $K'$ band. While \citet{deJong_1994} reported photometric errors (marked below with an asterisk), \cite{Tully_et_al_1996} did not. Below are shown the magnitudes for these galaxies in common:

\begin{center}
\begin{tabular}{c c c }
\hline
Galaxy & $m_{Ks}$ & $m_K$ \\
\hline
M100 & 6.50 $\pm$ 0.03 & 6.47 $\pm$ 0.23 (*) \\
NGC3726 & 7.61 $\pm$ 0.06 & 7.96 \\
NGC3877 & 7.68 $\pm$ 0.05 & 7.75 \\
NGC3938 & 7.79 $\pm$ 0.06 & 7.84 \\
NGC3953 & 7.00 $\pm$ 0.05 & 7.03 \\
NGC4013 & 7.58 $\pm$ 0.04 & 7.68 \\
NGC4157 & 7.31 $\pm$ 0.02 & 7.52 \\
\hline
\end{tabular}
\end{center}

Despite slight differences between the 2MASS $K_s$  and  the $K$-bands used in those previous studies, our measurements are in close  agreement with \cite{deJong_1994} and  \cite{Tully_et_al_1996}. As a further matter, if the errors reported by \cite{deJong_1994} are representative for early works, then the agreement arrived in this work with the studies of \cite{deJong_1994} and  \cite{Tully_et_al_1996} might be called remarkable.

We have also compared our results (Figures \ref{1-1_reff_OSU} and \ref{1-1_n_OSU}) with those from the Ohio State University Bright Spiral Galaxy Survey \citep[OSUBSGS,][]{Eskridge_2002} for the bulge parameters of effective radius and S\'ersic index in $H$ band reported by \citet{Laurikainen_et_al_2004_2}. We are in agreement within $1\,\sigma$ with the parameters reported by OSUBSGS.

\subsection{Bulge/Pseudobulge Classification}
\label{Class-Pseu_Res}

We implemented a bulge/pseudobulge classification using the following indicators:

\begin{itemize}

\item Fundamental Plane (FP) correlations are used for separating classical bulges from pseudobulges \citep[e.g.,][]{Kormendy-Ho_2013}. Hence, we adopted the Kormendy Relation \citep[KR,][]{Kormendy_1977}, a projection of the FP, as a tool to achieve this goal. Then, it is expected that classical bulges follow the KR, while pseudobulges are expected to fall as outliers of the relation \citep[e.g.,][]{Gadotti_2009, Anorve_PhD_Thesis}.

\item  Pseudobulges,  formed from secular processes, tend to have a rotational component and a low S\'ersic index $ n <2$ , while classical bulges have $n\geq2$ \citep{Fisher-Drory_2008}.

\item Classical bulges have high velocity dispersions, $\sigma \sim 160\; \mathrm{km\, s^{-1}}$  \citep{Fisher-Drory_2016}; while pseudobulges usually have $\sigma \sim  90\; \mathrm{km\, s^{-1}}$. 
Since the number of pseudobulges with
$\sigma > 130\; \mathrm{km\,s^{-1}}$ is low, we will call classical bulges those with $\sigma > 130\; \mathrm{km\, s^{-1}}$.

\end{itemize}

Therefore, {\em classical bulges must  hold  at least two of three indicators listed above}. It is worth mentioning that all of the galaxies' bulges (including those from E+S0 galaxies) have been classified. Therefore, we have that $\sim$60\% of the bulges in our sample are classical bulges, which is consistent in the three NIR bands.

\subsubsection{Kormendy Relation}
\label{KR_Res}

ETGs and some bulges display a link between the effective radius, $r_{e}$, and the surface brightness at the effective radius, $\mu_e$, known as the Kormendy Relation \citep[KR,][]{Kormendy_1977}, which implies that at the $r_e$, larger systems are fainter than the smaller ones, thus larger galaxies have lower densities. We have used distances of our sample to convert effective radii in physical units (as well as to compute luminosities for some scaling laws presented in the companion paper).
When we applied the KR as a tool to isolate classical bulges from pseudobulges, the latter ones will be those objects falling 3$\sigma$ out of the relation as outliers (see Figure \ref{KR_final}). From this analysis, we obtained that 71\% of the bulges are classical in $J$ band, while 70\% and 72\% in $H$ and $K_s$ bands, respectively.

Figure \ref{KR_final} shows the KR for the elliptical galaxies and bulges of our sample in the three bands of 2MASS, making a distinction among the sources classified as classical and pseudo bulges according to our scheme outlined above. E+S0 galaxies are closed in black circles, as well as dwarf ellipticals M110 and M32 that are also highlighted in black squares (see \S\ref{Ind_cases}). The linear fits in KR for each band (black line) were generated considering only E+S0 galaxies.

The linear fits to the KR for each band are given by the following expressions:  
\[
\begin{cases}
\mu_{e}(J) = (2.29 \pm 0.29) \logd r_{e}(J) + 16.54 \pm 0.33: & J \; {\rm band},\\
\mu_{e}(H)  = (2.35 \pm 0.24) \logd r_{e}(H) +  15.78 \pm  0.30: & H \; {\rm band},\\
\mu_{e}(K_s) = (2.55 \pm 0.27) \logd r_{e}(K_s) +  15.45 \pm  0.28: & K_s \; {\rm band}.
\end{cases}
\]

\subsubsection{S\'ersic index and velocity dispersion}

Based on the S\'ersic index criterion, we found that 60\%, 65\% and 61\% of the bulges are classical bulges, i. e., they have $n$ $\geq$ 2 in three NIR bands employed in this study.

Using the  velocity dispersion alone  as a classifier, we find that  51\% of the 97 objects with velocity data  in our  sample,  have
$\sigma \geq 130\; {\rm km\, s^{-1}}$.

\subsubsection{Final  Classical Bulge/Pseudobulge Classifications}

Thus, using the criteria established above,  we  have that  41\% ($J$), 42\% ($H$),  and 40\% ($K_s$) out of 101 galaxies in our  sample are pseudobulges. The last column of Table 3, which is presented in digital format, contains the assigned bulge classification for each galaxy.

\subsection{S\'ersic index, Bulge-to-Total and Axial Ratios}
\label{n-BT_Res}

The distributions of S\'ersic index, $n$,  are shown in Figure \ref{n_subplots_Galaxies} and Figure \ref{n_subplots_Spirals}. While, the distributions of Bulge-to-Total ratio, $B/T$, are shown in Figure \ref{BT_subplots_Galaxies} and Figure \ref{BT_subplots_Spirals}. These distributions are presented according to Hubble types and (sub)samples considered, such as the whole sample, E+S0 and spirals subsamples shown in panels $a$, $b$ and $c$ of Figures \ref{n_subplots_Galaxies} and \ref{BT_subplots_Galaxies}, respectively for the 2MASS bands; meanwhile, distributions for S, SAB and SB subsamples are shown in Figures \ref{n_subplots_Spirals} and \ref{BT_subplots_Spirals} for $J$, $H$ and $K_s$ bands in $a$, $b$ and $c$ panels, respectively. We also place these results in the context of bulge types. Also, the average values of $n$ and $B/T$ for these distributions are listed in Table \ref{Table_results_n_BT}. Besides, in Figure \ref{n-BT} we show the S\'ersic index as a function of $B/T$ ratio according to bulge types, where the criterion for $n$ used to classified bulges is also highlighted (dashed horizontal line). From these plots, the distinction between pseudobulges and classical bulges becomes more evident.

Furthermore, in Figure \ref{q_bulges} we present the axial ratio, $q$, distribution for the whole sample with the aim of looking for additional indicators on the morphology of bulge types. We will discuss later these findings that are consistent with earlier works.

\begin{table*}
\caption{\textbf{Distribution of $n$ and $B/T$.} Columns: (1) Group or subgroup of the sample: \textit{All} stands for the whole sample or subsample, while subscripts \textit{C} and \textit{P} for classical and pseudo bulges, respectively. (2)-(4) Mean S\'ersic index and (5)-(7) Mean $B/T$ in 2MASS bands.}
\label{Table_results_n_BT}
\begin{center}
\tabcolsep 5pt
\begin{tabular}{c c c c c c c}
\hline
Group  & $\mean n_J$ & $\mean n_H$ & $\langle n \rangle_{Ks}$ & $\mean {B/T}_J$ & $\mean {B/T}_H$ & $\mean {B/T}_{Ks}$ \\
(1) & (2) & (3) & (4) & (5) & (6) & (7) \\
\hline
All & 2.77 $\pm$ 0.15 & 2.82 $\pm$ 0.18 & 2.98 $\pm$ 0.17 & 0.36 $\pm$ 0.03 & 0.36 $\pm$ 0.02 & 0.38 $\pm$ 0.03 \\
All$_{C}$ & 3.49 $\pm$ 0.22 & 3.55 $\pm$ 0.22 & 3.66 $\pm$ 0.24 & 0.46 $\pm$ 0.03 & 0.45 $\pm$ 0.03 & 0.47 $\pm$ 0.03 \\
All$_{P}$ & 1.66 $\pm$ 0.15 & 1.72 $\pm$ 0.17 & 1.88 $\pm$ 0.25 & 0.21 $\pm$ 0.03 & 0.22 $\pm$ 0.03 & 0.25 $\pm$ 0.04 \\
E+S0 & 4.70 $\pm$ 0.46 & 4.57 $\pm$ 0.50 & 4.67 $\pm$ 0.48 & 0.64 $\pm$ 0.05 & 0.58 $\pm$ 0.06 & 0.62 $\pm$ 0.06 \\
E+S0$_{C}$ & 4.82 $\pm$ 0.46 & 4.66 $\pm$ 0.51 & 4.78 $\pm$ 0.50 & 0.64 $\pm$ 0.05 & 0.59 $\pm$ 0.07 & 0.63 $\pm$ 0.06 \\
Spirals & 2.28 $\pm$ 0.14 & 2.44 $\pm$ 0.14 & 2.55 $\pm$ 0.18 & 0.29 $\pm$ 0.02 & 0.30 $\pm$ 0.02 & 0.33 $\pm$ 0.03 \\
Spirals$_{C}$ & 2.87 $\pm$ 0.19 & 3.04 $\pm$ 0.17 & 3.17 $\pm$ 0.22 & 0.36 $\pm$ 0.03 & 0.38 $\pm$ 0.03 & 0.40 $\pm$ 0.03 \\
Spirals$_{P}$ & 1.64 $\pm$ 0.15 & 1.78 $\pm$ 0.18 & 1.86 $\pm$ 0.25 & 0.20 $\pm$ 0.03 & 0.22 $\pm$ 0.03 & 0.24 $\pm$ 0.04 \\
S & 2.51 $\pm$ 0.19 & 2.68 $\pm$ 0.20 & 2.70 $\pm$ 0.22 & 0.33 $\pm$ 0.04 & 0.34 $\pm$ 0.05 & 0.36 $\pm$ 0.05 \\
S$_{C}$ & 3.02 $\pm$ 0.18 & 3.22 $\pm$ 0.19 & 3.24 $\pm$ 0.17 & 0.39 $\pm$ 0.06 & 0.41 $\pm$ 0.05 & 0.41 $\pm$ 0.06 \\
S$_{P}$ & 1.92 $\pm$ 0.28 & 1.96 $\pm$ 0.29 & 1.85 $\pm$ 0.38 & 0.24 $\pm$ 0.07 & 0.24 $\pm$ 0.07 & 0.26 $\pm$ 0.08 \\
SAB & 2.36 $\pm$ 0.23 & 2.47 $\pm$ 0.22 & 2.46 $\pm$ 0.24 & 0.22 $\pm$ 0.03 & 0.23 $\pm$ 0.04 & 0.27 $\pm$ 0.04 \\
SAB$_{C}$ & 2.79 $\pm$ 0.32 & 2.92 $\pm$ 0.28 & 2.85 $\pm$ 0.28 & 0.28 $\pm$ 0.05 & 0.33 $\pm$ 0.05 & 0.37 $\pm$ 0.06 \\
SAB$_{P}$ & 1.82 $\pm$ 0.27 & 1.93 $\pm$ 0.29 & 2.05 $\pm$ 0.38 & 0.11 $\pm$ 0.02 & 0.12 $\pm$ 0.02 & 0.17 $\pm$ 0.04 \\
SB & 1.89 $\pm$ 0.30 & 2.07 $\pm$ 0.31 & 2.49 $\pm$ 0.50 & 0.34 $\pm$ 0.04 & 0.35 $\pm$ 0.04 & 0.37 $\pm$ 0.04 \\
SB$_{C}$ & 2.79 $\pm$ 0.49 & 2.95 $\pm$ 0.46 & 3.55 $\pm$ 0.78 & 0.42 $\pm$ 0.06 & 0.43 $\pm$ 0.06 & 0.45 $\pm$ 0.06 \\
SB$_{P}$ & 1.13 $\pm$ 0.13 & 1.46 $\pm$ 0.32 & 1.62 $\pm$ 0.54 & 0.28 $\pm$ 0.05 & 0.29 $\pm$ 0.05 & 0.31 $\pm$ 0.06 \\
\hline
\end{tabular}
\end{center}
\end{table*}

\section{Discussion}\label{Dis}

The 1D analysis presented by \citet{Schombert_2011} and \citet{Schombert_2012} reported the total magnitudes of ETGs in the 2MASS XSC \citep{Jarrett_et_al_2000} are systematically 0.33 mag fainter; such offset becomes stronger towards fainter magnitudes. However, in Figure \ref{1-1_mag_all} we do not find such effect. This can be explained by the fact that \citet{Schombert_2012} covered galaxies in a wider range in magnitude than us. \citet{Mezcua_et_al_2018}, also using  GALFIT to model the surface brightness of galaxies, reported the same discrepancy among their magnitudes with those from 2MASS XSC, which became larger for fainter galaxies. The source of this discrepancy lies on the generation of the LGA mosaics, which affected ETGs strongly. However, we have shown in Figure \ref{plot_SB_gals} that for LTGs, the LGA mosaics can be used safely.

With respect to the comparisons for S\'ersic index and effective radius for bulges shown in Figures \ref{1-1_reff_OSU} and \ref{1-1_n_OSU}, these results are consistent. We should point out that \citet{Laurikainen_et_al_2004_2} used a 2D modelling of surface brightness similarly to us. In addition, we are able to recover some well-established correlations that will be presented below.

As mentioned earlier, we have divided bulges in two different groups due to their properties that are related to different formation processes, though indications of different bulge populations were found observationally. Classical bulges have similar properties as elliptical galaxies, while the pseudobulges resemble discs, i.e., they have disc-like structure \citep[for reviews see][]{Kormendy-Kennicutt_2004, Kormendy-Ho_2013}. Then, it is expected that pseudobulges do not follow the same relations as E galaxies and classical bulges. As we said before, $\sim$60\% of the sample is classified as classical bulges. As the sample analysed in this paper is not complete in magnitude or volume, the distribution of bulges is somewhat biased. Nevertheless, since we have covered a wide galaxy morphology range (from E to SB galaxies), the properties of bulges and pseudobulges reported in this paper might be useful to anchor and compare with other galaxies samples at  low-$z$ and high-$z$, in the NIR and  other wavelengths. In particular, for large scale structure studies and the relation of Supermassive Black Holes with their host galaxies. 

Considering  the KR, one of the criteria used to separate bulge types, we found  agreement with previous studies \citep{Kormendy_1977, Pahre_et_al_1995, Anorve_PhD_Thesis, Olguin_et_al_2016}. Also, \citet{Schombert_2013} studied a sample of elliptical galaxies from 2MASS arriving at similar KR parameterisation. In addition, in Figure \ref{KR_final} most of the pseudobulges are falling outside of the correlation, which can be attributed to the distinct physical origin of pseudobulges that we outlined above. Additionally, we found that a significant amount of $n$ $<$ 2 fall as outliers in the KR: for $J$ band the 63\% pseudobulges according to KR have $n$ $<$ 2, while for the $H$ and $K_s$ bands, we found the 52\% and 68\%, respectively; this is in agreement with the results of  \citet{Gadotti_2009} and \citet{Anorve_PhD_Thesis}. Our results also show that some pseudobulges can be consistent with FP correlations and have larger effective radii and fainter surface brightness at the effective radius \citep{Kormendy-Ho_2013}, and some pseudobulges are  more compact than the classical bulges of the same luminosity \citep{Kormendy-Bender_2012}.

Another issue to highlight in our bulges classification is the presence of pseudobulges as a function of Hubble type, in the sense that pseudobulges are dominant towards LTGs (see Figures \ref{n_subplots_Galaxies}-\ref{BT_subplots_Spirals}). As we stated above, about 40\% of the objects in the entire sample are classified as pseudobulges, but when we remove the E+S0 galaxies from the whole sample, about 50\% of the spiral (S) galaxies host pseudobulges (see panel $c$ in Figures \ref{n_subplots_Galaxies} and \ref{BT_subplots_Galaxies}). Focusing on each spiral type and depending on the 2MASS band (see Figures \ref{n_subplots_Spirals} and \ref{BT_subplots_Spirals}), for S galaxies about 44\% of bulges are pseudo, while for SAB type about 51\% have pseudobulges and for barred galaxies the fraction of pseudobulges increases up to 56\%.

Also, from Figures \ref{n_subplots_Galaxies}-\ref{BT_subplots_Spirals} and Table \ref{Table_results_n_BT}, it can be seen that there is a correspondence between the S\'ersic index and Bulge-to-Total ($B/T$) ratio with Hubble types, since such parameters tend to be higher for ETGs (see panels $b$ in Figures \ref{n_subplots_Galaxies} and \ref{BT_subplots_Galaxies}). We also confirm the same trend for bulge types, i. e., classical bulges show higher values for $n$ and $B/T$ than pseudobulges. Furthermore, in our data it is evident a bimodality in such parameter distributions in the context of bulge types, especially in $n$, since the overlap is larger in $B/T$ distributions. This bimodality in $n$, as well as in $B/T$, has been noticed in previous results \citep[e.g.,][]{Allen_et_al_2006, Fisher-Drory_2008, Gadotti_2009}.

Thus, the average values of S\'ersic index when the complete sample (panel $a$ in Figure \ref{n_subplots_Galaxies}) is considered is about 2.8 (taking into account the three 2MASS bands for which results are very similar to each other). For the other subgroups like E+S0 and spiral galaxies (panels $b$ and $c$ in Figure \ref{n_subplots_Galaxies}, respectively), $\mean n \sim 4.6$ and 2.4, respectively. From here we see that the E+S0 galaxies tend to have higher S\'ersic indices than the whole sample and spiral galaxies (these last two groups tend to present similar values between them because most of the objects in our sample are spiral galaxies). These results agree with the expectation that ETGs tend to have high $n$ values (for instance, historically the shape of the light profiles for E galaxies are well described with the de Vaucouleurs function), while bulges in LTGs usually display low values of S\'ersic indices (bearing in mind that disc galaxies can be well modelled with an exponential profile). Besides, if we focus on the peaks of distributions, it can be noticed again that $n$ is higher for E+S0 galaxies, since the peak is $\sim$ 4 and higher, while in spiral galaxies and the whole sample the peaks are about 2.5, even though in $K_s$ band the peaks are at $\sim$1.5, which is in agreement with previous works where peaks of $n$ distributions range from 1-1.3 \citep[e.g.,][]{Blanton_et_al_2003, Anorve_PhD_Thesis}. Thus, we also confirm, as previous studies, the variation of S\'ersic index along the Hubble types of galaxies \citep[e.g.,][]{Graham_2001, Fisher-Drory_2008}. The same behaviour prevails when S galaxies are separated in groups, i. e., early-type systems in spirals, like the bulges of S galaxies, tend to show higher mean values and peaks in their distributions of $n$ than SAB and SB groups (see Figure \ref{n_subplots_Spirals}).

In the context of bulge types, it becomes clear from the plots how these two bulge populations display a division between them. When we consider the entire sample, the average values of $\mean n \sim 3.6$ for classical bulges and $\mean n \sim 1.8$ for pseudo bulges, respectively. For E+S0 galaxies, we have $\mean n \sim 4.8$ for classical bulges (only one E+S0 object was classified as pseudobulge, see \S \ref{Ind_cases}). For spiral galaxies, we report average values for $\mean n \sim 3$ for classical bulges and $\mean n \sim 1.7$ for pseudobulges, respectively. We also found that classical bulges have the following S\'ersic index distribution:   $\mean n \sim 3.2$ for  S galaxies, $\mean n \sim 2.8 $ for SAB galaxies, and $\mean n \sim 2.9$ for SB  galaxies,  respectively; while for pseudobulges the values of $n$ are distributed in the following manner: $\mean n \sim 1.9$ for S galaxies, $\mean n \sim 1.9$ for SAB galaxies, and $\mean n \sim  1.4$ for SB galaxies, respectively. Our results indicate that in all cases classical bulges have larger $n$ than pseudobulges. Also, we observe again a similar trend in the sense that classical bulges in ETGs exhibit higher S\'ersic indices than those from LTGs. The same trend prevails for the bulges of S, SAB and SB groups. For instance, \citet{Fisher-Drory_2008} found average $\mean n=3.49$ for classical bulges and $\mean n=1.69$ for pseudobulges for a sample of galaxies ranging in Hubble type from S0 to Sc, which also compares in size with ours. Besides, when they add elliptical galaxies to the classical bulges subsample, the value incresas to $\mean n  \sim 3.78$.

Regarding the distributions of $B/T$, we can see also how it varies as a function of the Hubble type as the case for $n$, where ETGs tend to have higher values than those in LTGs (see Figures \ref{BT_subplots_Galaxies} and \ref{BT_subplots_Spirals}). This drop in $B/T$ values is consistent with previous works \citep[e.g.,][]{Graham_2001, Laurikainen_et_al_2004, 2017A&A...598A..32M}. Analogously to S\'ersic index, $B/T$ distribution is bimodal despite the overlap is larger than in the case of $n$. The average value of $B/T$, considering the three bands, is $\mean {B/T} \sim 0.37$ for the complete sample; while the E+S0 and spirals subgroups show mean values of $\mean {B/T}  \sim {0.61}$ and $\mean {B/T} \sim {0.30}$, respectively (see panels $a$, $b$ and $c$ in Figure \ref{BT_subplots_Galaxies} for each galaxy group). If we see the subsamples of spiral galaxies  separately for each 2MASS band (see panels $a$, $b$ and $c$ in Figure \ref{BT_subplots_Spirals}), then $\mean {B/T} \sim 0.34$ for S galaxies, $\mean {B/T}  \sim 0.24$ for SAB galaxies and $\mean {B/T} \sim 0.34$ for SB galaxies, respectively.

For pseudobulges in the whole sample, these tend to have lower values of $\mean{B/T} \sim {0.23}$, while classical bulges have $\mean {B/T} \sim 0.46$. The classical bulges of E+S0 galaxies have $\mean {B/T} \sim 0.62$ (again, we recall that there is one object classified as pseudobulge in the E+S0 subsample, see \S \ref{Ind_cases}). For the group of spirals, the average value of $\mean {B/T} \sim 0.38$ for classical bulges, while for pseudobulges is $\mean {B/T} \sim 0.22$. Again, we see that classical bulges display higher $B/T$ values than pseudobulges. The same trend is observed in the bulges of S, SAB and SB galaxies. Thus, our findings are in good agreement with previous ones reported, such as those values from \citet{Fisher-Drory_2008} of $\mean {B/T}  = 0.41$ for classical bulges and 0.16 for pseudobulges. Besides, similarly to \citet{Gadotti_2009}, the peaks in our distributions for pseudobulges is $\sim$ 0.10, while for classical bulges is $\sim$ 0.50 (see Figure \ref{BT_subplots_Galaxies}).

Our results show that most of pseudobulges have low values of $B/T$, while classical bulges cover almost the entire range of $B/T$ distribution. This result is consistent with the one of \citet{Fisher-Drory_2009}, who reported that pseudobulges are more likely to be found in low $B/T$ galaxies. Our results also show a large overlap in $B/T$ distribution of bulges and pseudobulges, confirming that a low $B/T$ value does not secure that a galaxy hosts a pseudobulge, and that if a bulge has a $B/T$ $\geq$ 0.5, then it can be expected to be classical \citep[e.g.,][]{Kormendy-Kennicutt_2004,Kormendy-Ho_2013}. Besides, even if some pseudobulges can have $B/T > 0.2$ \citep{Fisher-Drory_2008}, in our data it is visible how the number of pseudobulges decreases towards $B/T\sim 0.5$. A similar behaviour is shown in the distribution of $B/T$ by \citet{2020arXiv200507588P}, who compiled a sample of mostly disc galaxies from literature \citep{Kormendy_et_al_2010, Fisher-Drory_2011}. Furthermore, for the galaxies in common with \cite{2020arXiv200507588P} we performed a Kolmogorov-Smirnov test and obtained a statistic of 0.35 at the 95\% significance level, hence, we retain the null hypothesis and conclude that both $B/T$ data sets are drawn from the same parent distribution. In fact, from the $\sim$ 40 pseudobulges in our sample in the three bands of 2MASS, we found that about 12\% of them have $B/T \geq 0.5$. Additionally, we find a good agreement with theoretical predictions of semi-analitycal models by \citet{Izquierdo_2019}, who show some bulge properties such as $B/T$ for both populations of bulges. The distribution of pseudobulges agrees with ours, although they do not fully reproduce the peak towards $B/T \sim$ 0.1. Nonetheless, in the case of classical bulges differences arise, since the $B/T$ distribution peaks about 0.1 and abruptly decreases towards values close to 0.5, which, as we said above, is an observed threshold for delimiting bulge types that our results also confirm. \citet{Izquierdo_2019} argue that this effect is because their model produces classical bulges in galaxies with too-massive stellar discs.

The relation between $B/T$ and S\'ersic index in Figure \ref{n-BT} shows that most of classical bulges tend to locate in the upper region of these plots, presenting larger values for $n$ and $B/T$. On the other hand, pseudobulges tend to present lower values for these parameters. Also, we can see some pseudobulges with high values of $B/T$ and $n\geq2$  and, conversely, some classical bulges with low values of $B/T$ and $n<2$. This result confirms what we have said before about the distribution for these parameters and is in agreement with previous works \citep[e.g.,][]{Laurikainen_et_al_2004, Fisher-Drory_2008, Weinzirl_et_al_2009, Gadotti_2009}. Nevertheless, a correlation between S\'ersic index and $B/T$ is weak at the most, especially for pseudobulges. We further discuss the properties and tendencies  displayed by pseudobulges, classical bulges and elliptical galaxies in the companion paper. 

Therefore, we can conclude that our results also show a bimodality present in the distributions for $n$ and, to a lesser extent, for $B/T$. Figure \ref{n-BT} shows that classical bulges and pseudobulges cannot be distinguished using  $B/T$ alone, and that the $n\geq2$ threshold provides a better separation. Thus, contrary to \citet{Kormendy-Ho_2013}, we state that $B/T$ cannot be used as a single bulge classifier in the NIR.

Similarly, the distribution of the  axial ratio $q$  for bulges does not display a clear separation between pseudobulges and classical bulges, in fact both distributions look practically indistinguishable showing a very extended overlap in almost all the dynamic range (see Figure \ref{q_bulges}). The peaks of distributions are located at $q$ $\sim$ 0.8 in the three bands. This result is qualitatively in agreement with previous works \citep[e.g.,][]{Padilla_2008, Bruce_et_al_2012}.

According to \citet{Kormendy-Kennicutt_2004}, the morphology of bulges can be used as a criterion to distinguish between bulges, since pseudobulges tend to have a disc-like morphology, i.e., they show an apparent flattening similar to that of the outer disc. Thus, we would expect to see more pseudobulges with lower $q$ than classical bulges. However, this tendency is barely seen in Figure \ref{q_bulges} in the low amplitude tails when $q$ tends to zero.

\subsection{Notes on Individual Galaxies}
\label{Ind_cases}

We will discuss below some cases related to their bulge classifications. Firstly, we consider the galaxy  M110 that  has been classified as pseudobulge following the criteria of KR and low velocity dispersion, in fact, it is an extreme outlier of the KR and the only ETG classified as pseudobulge (see Figure \ref{KR_final}). This object, is a dwarf elliptical galaxy classified also as peculiar \citep[E5,pec;][]{deVaucouleurs_et_al_1991}. M110, as well as M32, are  satellites of M31. Evidence suggests that substantial stellar material (found in the halo of M31) has been stripped from these satellite galaxies after tidal interactions with M31 \citep{Ibata_2001}. As a matter of fact, \citet{Jarrett_et_al_2003} also pointed out the particular case of M110, arguing that its radial surface brightness is dominated by an exponential disc, which reinforces the issue that M110 was fitted with two components (see \S \ref{E_gals_double}). Regarding the compact elliptical galaxy M32 (cE2), it is classified as classical bulge because of the KR and $n$ criteria, and, similarly to M110, it was modelled with two components [an exponential disc as an additional component; actually, \citet{Graham_2002} reported that its surface brightness distribution is well described by a bulge plus an exponential disc profile]. Besides, it has been proposed that it was disc galaxy with initial  luminosity close to that of our Galaxy, which was disrupted due to an  encounter with Andromeda galaxy \citep{DSouza_2018}. So, this may be the reason why M32 is falling almost in the $3\sigma$ boundary of the KR (see Figure \ref{KR_final}). Hence, the low values in surface brightness, S\'ersic index and velocity dispersion of M110, as well as the case of M32, can be a consequence of  tidal interactions with M31.

We have a few cases (only 4$\%$ of the sample) where the bulge classification does not arrive at the same bulge type in the three NIR bands. The galaxies with discrepant classification are the following: NGC4710, NGC3344, M51b and NGC7582. We decided to assign as the final classification the one that was common in the two other bands. Thus, the bulge of NGC4710 is classified as classical in $H$ and $K_s$ bands, while the one of NGC3344 is classified as pseudobulge only in $H$ band. On the other hand, the bulge of M51b is a pseudobulge in bands $J$ and $H$, while the bulge of NGC7582 is classified as pseudobulge in $H$ and $K_s$ bands.

In addition, the galaxies M83 and NGC5792, lacking of reported velocity dispersion data, do not fit into either of the two bulge categories. These galaxies have $n<2$ and do follow the KR, but to assign a classification, we considered less established properties of pseudobulges, such as $B/T$ ratio and the presence of bars within the bulge region \citep{Fisher-Drory_2008}, since M83 (SAB) and NGC5792 (SB) are barred galaxies and have $B/T \lesssim 0.20$. Hence, our final classification is that the bulges in M83 and NGC5792 are pseudobulges.

We have also compared our bulge classifications with some previous works. For instance, we have 10 disc galaxies in common with \citet{Kormendy-Ho_2013}, six of them are  classical bulges, the other four are  pseudobulges. We agree on the classical bulges, however, we disagree on the classification of NGC4826 and NGC4945 as pseudobulges. \citet{Kormendy-Ho_2013} used more classification criteria than us \citep[the list was introduced by][and an object must hold at least two of them]{Kormendy-Kennicutt_2004}. After inspecting such list given also by \cite{Kormendy_et_al_2011}, we noticed that the bulge of NGC4826 was classified as a pseudobulge because the center of the galaxy is dominated by Population I material and there is no sign of a merger in progress, besides NGC4826 is rotation-dominated and is a low-$\sigma$ outlier in the Faber-Jackson Relation (FJR) according to \citet{Kormendy_et_al_2011}. However, for us NGC4826 follows the KR, it has $n> 2$ and  $\sigma=96\; {\rm km\, s^{-1}}$ (in fact, they also report such $\sigma$ value). Nevertheless, in a plot of the FJR (Paper II), it can be seen that it is not a low-$\sigma$ outlier. As we see, velocity dispersion information is the only point in common we do agree with, so a possible explanation for this discrepancy may be related to the difference in the surface brightness modelling, since 1D decomposition was used by \cite{Kormendy-Ho_2013}, unlike the 2D approach presented in this paper.

Regarding NGC4945, a barred galaxy viewed edge-on with a dusty bulge, we failed to find the criteria by which it was classified as pseudobulge. However, it is also considered as a pure-disc galaxy viewed edge-on that contains neither a classical nor a pseudobulge \citep{Kormendy_et_al_2011}. Nevertheless, we found that the bulge of NGC4945 is more prominent in the $K_s$ band. Then, we have classified NGC4945 as classical bulge due to its $n > 2$ and $\sigma > 130 \; {\rm km\, s^{-1}}$, in spite of being an outlier in the KR.

Also, when comparing with the identification of bulge types performed by \citet{Fisher-Drory_2008}, we found that from the 16 galaxies in common, only 3 bulges are classical and the rest are labelled as pseudobulges. We agreed on the 3 classical bulge classifcations, but disagree on 7 of the pseudobulges. We disagree on the pseudobulge type assigned to the bulges of the galaxies NGC4826, NGC3166, NGC4569, NGC4579, M63, M88 and M106 [the latter one is classified as a classical bulge by us, in agreement with \citet{Kormendy-Ho_2013}]. \citet{Fisher-Drory_2008} classifications are based on the morphology of the galaxy in the bulge region, which is the only classification criterion used. If the bulge contains a nuclear bar, a nuclear spiral and/or a nuclear ring, it is classified as a pseudobulge by \citet{Fisher-Drory_2008}. Nonetheless, as has been shown  in previous studies \citep[e.g.,][]{Graham_2008, Kormendy-Ho_2013} and in this paper, barred galaxies can host classical bulges. This supports the view that using a single criterion as suggested by \citet{Fisher-Drory_2008} may not be sufficient to classify bulges  in barred galaxies \citep[e.g.,][]{2021MNRAS.502.2446E}.

The disagreements discussed here suggest that a more detailed inspection for these galaxies may be required, since they may contain peculiar bulges, which in turn indicates that our classification scheme may be incomplete. Thus, we should mention that in this paper we are currently unable to classify bulges that may be composite systems, with both pseudobulge and classical bulge features, that may co-exist in the bulges of some galaxies \citep{Kormendy-Kennicutt_2004, Kormendy-Ho_2013}; for example the bulge of NGC4826 was classified as a mixed bulge by \citet{Kormendy_et_al_2010}. Complementary dynamical studies could be useful to explain the nature of composite pseudo-classical bulges \citep[e.g.,][presented a very complete analysis combining photometry with IFU stellar kinematics]{2021MNRAS.502.2446E}. Such studies may help us to establish more robust criteria for recognising between different types of bulges.

\section{Summary and Conclusions}\label{Con}

In this work we studied a sample of bright and nearby galaxies by modelling their surface brightness through a two-dimensional photometric decomposition in order to obtain information about their structural parameters. This sample of galaxies spans from early to late-type morphologies. We remind  the reader that measured parameters from our 2D photometric decomposition carried out with GALFIT are presented in tables available only in digital format. We have compared our results obtained  with previous works using a 1D technique and we found consistency in them. Besides, we have looked into the issue related to the 2MASS galaxy photometry. Also, from a joint analysis between structural parameters, scaling relations and kinematic measurements (the latter information taken from literature), we classified bulges in galaxies into  classical bulges and pseudobulges. Below we summarise our main conclusions:

\begin{enumerate}

\item We performed a 2D surface brightness modelling for the galaxies in our sample using GALFIT. The sample comprises 101 bright and nearby galaxies observed by 2MASS survey in the NIR bands. From this analysis, we obtained information for the structural parameters of galaxies, such as concentration indices, effective radii, magnitudes, axis ratios, among others.

\item The 2D photometric decomposition carried out in this work has been more detailed than other works, in the sense that many of them at most include a bulge+disc decomposition, while our models generated with GALFIT go further by considering a bar and an additional component in some cases. Similarly, many of the elliptical galaxies in our sample are described by models involving a S\'ersic component plus an additional component. We show that in the NIR the inner component suggested by \citet{2013ApJ...766...47H} is absent. This might be related to the difference in stellar populations that are sampled in optical which differ in the  NIR.

\item We have addressed the sky over-subtraction issue related to the 2MASS data for the LGA mosaics, where the outer region of large ETGs was removed by using an aperture that was too close to the galaxy during the mosaics construction. Alternatively, we used the IRSA tile service to generate images to perform comparisons between both data sets, i.e., LGA mosaics and image tiles. In \S \ref{LGA_VS_tile} we demonstrated that the over-subtraction problem present in the LGA mosaics affected ETGs strongly, while LTGs were unaffected. The total magnitudes measured on LGA mosaics presented a mean offsets of 0.32 mag, 0.34 mag, and 0.34 mag for $J$, $H$ and $K_s$, respectively. Therefore, we propose to use image tiles generated through IRSA server for a more accurate appraisal of the surface brightness photometry for ETGs, instead of using LGA mosaics.

\item We have compared our results with published 1D photometric analyses, finding close agreement. Nevertheless, our 2D approach considers more degrees of freedom allowing to disentangle the contributions of different galaxy components, such as bulge, disc and bar, in a more reliable fashion. This also allows us to recover some well-established correlations between structural parameters of galaxies.

\item As it has been suggested in many studies \citep[e.g.,][]{Kormendy-Ho_2013, Fisher-Drory_2016}, the use of only one criterion alone to identify bulge varieties does not provide a robust classification. Therefore, we use the  following  criteria: the Kormendy Relation, S\'ersic index and velocity dispersion. We consider as pseudobulges those points falling as outliers in Kormendy Relation. Also a low S\'ersic index, $n < 2$, could be an indicator of pseudobulge; otherwise if $n \geq 2$, it is considered as a classical bulge. Regarding the velocity dispersion, if a bulge has a velocity dispersion 
$< 130 \; {\rm km\,s^{-1}}$, it is likely to be a pseudobulge, while above this threshold it would be regarded as a classical bulge. Hence, if one object holds at least two of these conditions, it is classified as classical bulge or pseudobulge. Thus, we obtained that $\sim 40 \%$ of the sample is classified as pseudobulges in the three bands of 2MASS.

\item Our results confirm the previous distributions for some properties of classical bulges and pseudobulges, such as S\'ersic index, $B/T$ ratio as well as $q$ ratio \citep[e.g.,][]{Fisher-Drory_2008, Kormendy-Ho_2013}. The bimodality is more evident for $n$, while for $B/T$ ratio the overlap between classical and pseudo bulges is large. For this reason, we did not use the $B/T$ ratio as an auxiliary criterion for bulge classification. Our results indicate that classical bulges and ETGs tend to have higher values of $n$ and $B/T$ than pseudobulges and LTGs. Similarly, in the axial ratio distribution is evident even a more extended overlap between bulge types in almost all the dynamic range, only towards low values of $q$ ratio we can marginally see a trend in which the presence of pseudobulges increases. We found, that to a greater extent, pseudobulges tend to be found in LTGs, since the fraction of pseudobulges increases from S to SB galaxies.

\item We also made a direct comparison of our bulge classification for  objects in common  with previous studies \citep{Fisher-Drory_2008, Kormendy-Ho_2013}. In general, we agreed on the classifications of classical bulges, but disagreed in some cases that were classified as pseudobulges. The classification scheme advanced in this paper does not include composite systems, i.e., those bulges  sharing classical/pseudo bulges features. Thus, a more refined classification might need to include IFU complementary kinematic information.

In a companion paper, we explore some of the most common galaxy scaling relations for the sample presented in this paper, as well as correlations between the mass of SMBH and global properties of the galaxies.

\end{enumerate}

\section{Data Availability}

The data underlying this paper are available on tables and  online supplementary material. Other datasets were derived from sources in the public domain: 2MASS Large Galaxy Atlas at \url{https://irsa.ipac.caltech.edu/applications/2MASS/LGA/}, 2MASS image tiles at \url{https://irsa.ipac.caltech.edu/applications/2MASS/IM/interactive.html#pos}, NASA Extragalactic Database at \url{http://ned.ipac.caltech.edu/} and Hyperleda database at \url{http://leda.univ-lyon1.fr/leda/param/vdis.html}.

\section{Acknowledgements}
We are very grateful to the anonymous referee for very constructive comments and for pointing out the background  over-subtraction problem on 2MASS LGA mosaics. 
ER-L is grateful to Mexico's National Council of Science and Technology (CONACyT) for supporting this research under PhD studentship grant and to Elena Terlevich for the SNI-CONACyT graduate assistantship. JA-E, MV and GI are grateful to {\em Verano de la Investigaci\'on Cient{\'\i}fica en el INAOE (VICI)}, an INAOE's Summer Research Program sponsored by the Mexican Academy of Science (AMC), Programa DELFIN, CONACyT, and INAOE. We wholeheartedly thank Prof. Tom Jarrett for sharing 2MASS reduced, star-removed, data cubes and for all  the useful discussions that helped to improve this paper. We  thank Prof. James Schombert for sharing data and for the useful exchange of ideas. We also thank Ramondavid R{\'\i}os for proofreading the article. Last but not least, we want express our most gratitude to Chien Peng for his constant, helpful, and friendly  advice on the use of GALFIT, all along the different stages of our project.

This publication makes use of data products from the Two Micron All Sky Survey (2MASS), which is a joint project of the University of Massachusetts and the Infrared Processing and Analysis Center/California Institute of Technology, funded by the National Aeronautics and Space Administration and the National Science Foundation. We also acknowledge the use of the NASA/IPAC Extragalactic Database (NED) and HyperLeda databases.


\bibliography{Bib.bib}

\appendix\label{Append}
\section{Estimation of  Uncertainties for GALFIT Parameters}

Previous works have shown that GALFIT underestimates errors \citep[e.g.,][]{Haussler_et_al_2007, Tarsitano_2018}. Here, we treat this problem through the model fitting of artificial galaxies. In other words, we have created galaxy images which resemble the 2MASS galaxy images in our sample and then we used GALFIT to fit those galaxies to estimate the parameter errors.

An important step to build mock galaxies is to model errors as accurately as possible and identify model parameters that may contribute significantly to errors. Among those main components we have atmosphere blurring, Poisson error, sky error and object contamination by nearby objects, such as other galaxies or stars.

As it can be seen in the images, 2MASS galaxies for our sample are brighter than any other object within the same image. This means that objects such as stars (or other galaxies) make little contribution to the errors. This implies that we can construct reliable isolated artificial galaxies since the main galaxy is practically unaffected by other objects of the image. In addition, as it was explained earlier, we use SExtractor to make masks for undesired objects in the image.

Another minor error source to surface brightness modelling is atmosphere blurring. In our 2MASS sample, galaxies are larger than the average PSF (e.g. $\sim$3.2'' for $J$ band) which implies that the surface brightness model is almost unaffected by PSF's convolution. We have tested this by removing the PSF models during the fits and, as a result, the differences between the model with and without PSFs are minimum ($\sim$0.01 difference in S\'ersic index and other cases nearly $\sim$1). However, as we stated above, we remind  the reader that the models generated with GALFIT for the galaxies in our sample are convolved with the PSF image.

We also take into account the error contribution by the sky background, since it may slightly influence the model parameters despite we use observations in NIR bands. In particular, bands $H$ and $K_s$ reach values of 1 in $\sigma$ dispersion (see Table \ref{Table_means_of_Sky_and_PSFs}). This becomes the main source of errors for our fittings. Below, we explain how we construct and fit artificial galaxies.

\subsection{Generation of Artificial Galaxies}

We have made isolated galaxies on images of size 500 $\times$ 500 pixels for each band of the 2MASS data. This is the smallest size for our 2MASS's images. We have created $1000$ galaxy images for each band $J$, $H$ and $K_s$ (i.e., 3000 in total). To construct galaxies, we have modelled them from surface brightness models. As in the fitting procedure for our 2MASS data, S\'ersic profile is used for single component galaxies, while S\'ersic and exponential models are used for bulge and disc, respectively.

\begin{figure}
\includegraphics[width=8.45 cm]{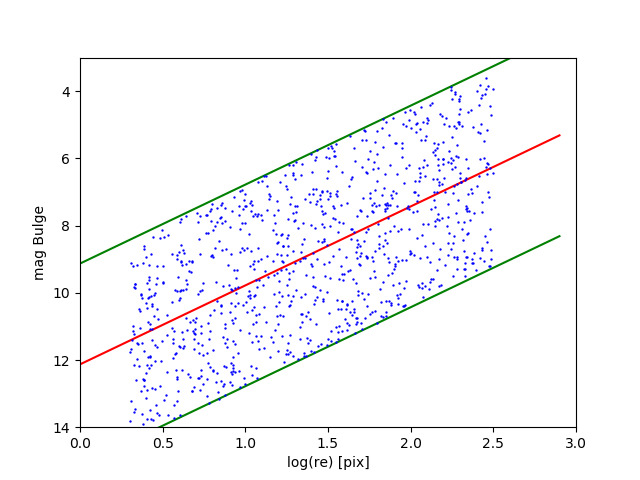}
\caption[mag-re]{\footnotesize Magnitude-effective radius relation for 1000 simulated galaxies. Y-axis represent apparent magnitude and X-axis represent effective radius in pixels. Blue points represent artificial bulges. Red solid line is the liner fit to the relation taken from our data in $J$ band. Green solid lines represents the limits for the parameters of our artificial galaxies.}
\label{mag-re}
\end{figure}

We have obtained the model parameters from the magnitude-effective radius relation result of the simulation process in a similar way as \citet{Haussler_et_al_2007}. For instance, for one galaxy we have selected a random $\log(r_e)$ (in pixels) and computed the magnitude at the effective radius according to the magnitude-effective radius relation that we have obtained from our sample. In Figure \ref{mag-re} is shown an example of the bulge distribution obtained from our data in $J$ band.

Tables \ref{Table_range_of_pars1} and \ref{Table_range_of_pars2} summarise the range of parameters used for both S\'ersic and Bulge+Disc artificial galaxies, respectively. Unlike \citet{Haussler_et_al_2007}, we treat the S\'ersic index $n$ and sky background as free parameters.

\begin{table}
\caption{Range of parameters used for artificial galaxies (S\'ersic component)}
\label{Table_range_of_pars1}
\begin{center}
\tabcolsep 4pt
\begin{tabular}{ l r r }
\hline
Parameter& Min &  Max \\
\hline
Flux(mag) & 1.8 &  11.6 \\
Effective radius ($\logd r_{e}$)& 0.3 &   2.5 \\
S\'ersic index ($n$) & 0.5 &  7.0 \\
Axis Ratio ($q$) & 0.4 &  1.0 \\
Position Angle ($\theta$) &-90.0 &  90.0 \\
\hline
\end{tabular}
\end{center}
\end{table}

\begin{table}
\caption{Range of parameters used for artificial galaxies (Bulge+Disc components)}
\label{Table_range_of_pars2}
\begin{center}
\tabcolsep 3pt
\begin{tabular}{ l r r r l r r }
\hline
\multicolumn{3}{c}{\bf{Bulge}} & \multicolumn{4}{c}{\bf{Disc}}\\
Parameter & Min &  Max & \vline & Parameter & Min &  Max \\
\hline
mag & 3.3 &  14.1 & \vline &mag & 0.6 &  12.7 \\
$\log r_e$&  0.3  & 2.5 & \vline & $\log r_s$ &1.0 &  3.0 \\
$n$ &  0.5 & 7.0 & \vline & $\cdots$  & $\cdots$ & $\cdots$ \\
 $q$ &0.4 &   1.0 & \vline & $q$&  0.1 & 1.0 \\
$\theta$ & -90.0 &  90.0 & \vline & $\theta$ & -90.0 &  90.0 \\
\hline
\end{tabular}
\end{center}
\end{table}

Once we obtained the catalogue file, we proceeded to construct galaxies using GALFIT. The reliability of GALFIT for constructing S\'ersic models has been tested before \citep{Haussler_et_al_2007}. Also, here we focus on testing how the sky noise affects our fits.

We have made a file with the initial parameters for every galaxy in the catalogue. Then, we proceeded to run GALFIT on every file to obtain a galaxy image. This image does not contain any source of noise.

To insert the sky noise, we first estimate the mean and standard deviation in separated regions of the sky (near to the corners; see Table \ref{Table_means_of_Sky_and_PSFs}) of every 2MASS image. We have selected squares with sizes of 200 x 200 pixels in empty regions.
 
\begin{table}
\caption{Sky mean values for the sky background and PSF's FWHM}
\label{Table_means_of_Sky_and_PSFs}
\begin{center}
\tabcolsep 4pt
\begin{tabular}{ c c c r c}
\hline
\multicolumn{3}{c}{\bf{Sky}} & \multicolumn{2}{c}{\bf{PSF}}\\
Band & $\mu$ & $\sigma$ & \vline & $\mean {FWHM}$ \\
\hline
$J$ & 0.007 & 0.60 & \vline & 3.20 \\
$H$ & 0.002 & 1.00 & \vline  & 3.10 \\
$K_s$ & 0.008 & 1.02 & \vline  & 3.14 \\
\hline
\end{tabular}
\end{center}
\end{table}

Using the means of the PSF's FWHM per each band listed in Table \ref{Table_means_of_Sky_and_PSFs}, gaussian models were used to replicate the PSF used for convolving with the galaxy model. Then, Poisson noise was also added. As a final step, we have added a template with sky background to this last image.

Following, we proceeded to fit the galaxies using GALFIT. We used a small script to automate this process. This script runs SExtractor on each image and adapts its output catalogue to create GALFIT initial parameter files. From the 1000 galaxies in each band, GALFIT was able to fit above the 90 $\%$ of the galaxies in each band for every case of single, bulge and disc components.

For the case of single S\'ersic galaxies, we used 2 components to fit: a S\'ersic model and sky model. On the other hand, for Bulge+Disc galaxies we used 3 components: S\'ersic, Exponential disc and sky model.

Sky component was regarded as a free parameter during the fit (as we did with fittings on 2MASS images). Once we have the object file, the script creates a list file which contains the file path of the initial parameter files to run with GALFIT. Finally, we run this list file along with GALFIT.

When GALFIT has finished, we compare the fitted model galaxies with their true model values to estimate the parameter errors. The results are shown in the next section.

\subsection{Resulting Uncertainties}

Here we show the results of the fitting of the simulated galaxies created as it was explained above. In Tables \ref{Table_results_Sersic} and \ref{Table_results_BD} are summarised the results for galaxies with a single S\'ersic and Bulge+Disc components, respectively. The columns show the deviations from the simulated value and scatter for the most important fitting parameters as magnitude, $r_e$ (in pixels) and $n$. The mean and $\sigma$-values given are computed after applying a 3$\sigma$ clipping to those simulated galaxies that were successfully fitted by GALFIT. Hence, the number of galaxies was $>$ 80\% of the 1000 simulated galaxies in most of the parameters computed. In fact, it can be seen how sky errors (see Table \ref{Table_means_of_Sky_and_PSFs}) are related to uncertainties reported in Table \ref{Table_results_Sersic}, in the sense that sky dispersion in the $K_s$ band tend to be higher and as  a result,   the uncertainties for parameters such as magnitude, $r_e$ and $n$ are also  higher.

\begin{figure} 
\begin{center}
\includegraphics[width=8.25 cm]{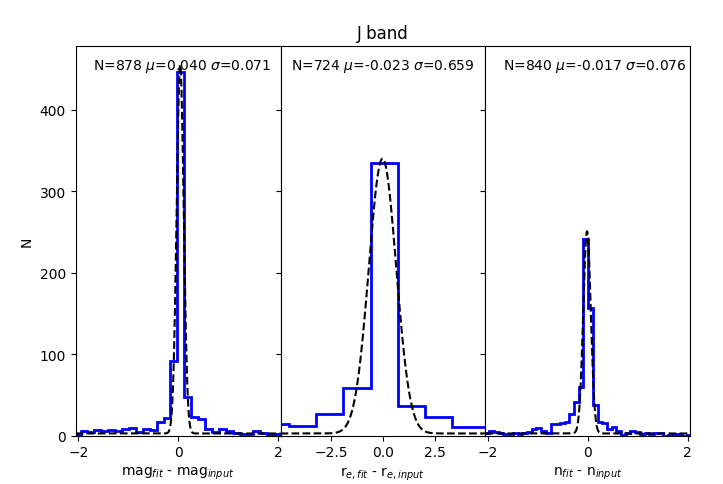}
\caption[uncertainties_J]{\footnotesize Histograms showing the deviations for the fitting parameters analysed in the simulations, $\Delta$mag, $\Delta$r$_e$ and $\Delta$n, generated from data in the $J$ band for bulges.}
\label{uncertainties_J}
\end{center}
\end{figure}

As it can be noticed in Figure \ref{uncertainties_J}, when we computed the differences between the simulated and the fitted galaxies for these three parameters, there are no significant deviations from the expected values, where a really good parameter recovery can be seen. This agreement is due to the high S/N in the observed galaxies, which is secured by the high flux of the galaxies in our sample ($K_s < 10$, see Figure \ref{Hist-Hubble_type}), the high resolution achieved on each galaxy due to their large size, which make the effects of seeing almost negligible,  and the relative low deviations in the background that is indicated by the tight parameter recovery. This, however, is not seen in other works when fainter galaxies are considered \citep[e.g.,][]{Haussler_et_al_2007, Anorve_PhD_Thesis}. Hence, final error bars used in the analysis of this work are obtained by adding in quadrature the uncertainties reported in this appendix (Tables \ref{Table_results_Sersic} and \ref{Table_results_BD}) with the ones from GALFIT models reported in Tables 3 to 6.

\begin{table}
\caption{Results from simulations for galaxies with a single S\'ersic component}
\label{Table_results_Sersic}
\begin{center}
\tabcolsep 4pt
\begin{tabular}{ c c c c c c c c c}
\hline
\multicolumn{1}{c}{} & \multicolumn{2}{c}{$\Delta$ mag} & \multicolumn{3}{c}{$\Delta$ $r_e$} &  \multicolumn{3}{c}{$\Delta$ $n$} \\ \cline{2 - 3} \cline{5 - 6} \cline{8 - 9}
Band  & Mean & $\sigma$ & & Mean & $\sigma$ & & Mean & $\sigma$ \\
\hline
 $J$ & -0.002 & 0.008 & & 0.037 & 0.272 & & -0.002 & 0.007 \\
 $H$ & 0.006 & 0.008 & & 0.211 & 0.348 & & 0.002 & 0.010 \\
 $K_s$ & -0.005 & 0.068 & & -0.210 & 0.475 & & -0.001 & 0.018 \\
\hline
\end{tabular}
\end{center}
\end{table}

\begin{table}
\caption{Results from simulations for galaxies with bulge and disc components}
\label{Table_results_BD}
\begin{center}
\resizebox{\columnwidth}{!}{
\tabcolsep 2 pt
\begin{tabular}{ c c c c c c c c c r c c c c c }
\hline
\multicolumn{8}{c}{\bf{Bulge}} & \multicolumn{7}{c}{\bf{Disc}} \\
\multicolumn{1}{c}{} & \multicolumn{2}{c}{$\Delta$ mag} & \multicolumn{3}{c}{$\Delta$ $r_e$} & \multicolumn{3}{c}{$\Delta$ $n$} & \multicolumn{3}{c}{$\Delta$ mag} & \multicolumn{3}{c}{$\Delta$ $r_s$} \\ \cline{2 - 3} \cline{5 - 6} \cline{8 - 9} \cline{11 - 12} \cline{14 - 15}
Band & Mean & $\sigma$ & & Mean & $\sigma$ & & Mean & $\sigma$ &  & Mean & $\sigma$ & & Mean & $\sigma$ \\
\hline
 $J$ & 0.040 & 0.071 & & -0.023 & 0.659 & & -0.017 & 0.076 & \vline & 0.006 & 0.060 & & -0.145 & 0.612 \\
 $H$ & 0.030 & 0.056 & & 0.061 & 0.556 & & -0.005 & 0.049 & \vline & 0.018 & 0.075 & & 0.062 & 0.618 \\
 $K_s$ & 0.010 & 0.019 & & 0.605 & 0.757 & & -0.007 & 0.063 & \vline & 0.020 & 0.059 & & -0.082 & 0.538 \\
\hline
\end{tabular}}
\end{center}
\end{table}

\end{document}